\documentclass[11pt]{article}
\pdfoutput=1
\usepackage{amsmath,amssymb,epsfig,amsfonts,amsthm}
\usepackage{graphicx,subfigure}
\usepackage[usenames, dvipsnames]{color}
\usepackage{mdframed}
\usepackage{hyperref}
\usepackage{cite}
\usepackage{verbatim} 
\usepackage{float}
\restylefloat{table}
\usepackage[all]{xy}
\usepackage{pdflscape}
\usepackage{array}
\usepackage{youngtab}
\usepackage{multirow}
\usepackage{color}
\usepackage{ulem}

\makeatletter

\def\wL{\mathbb{L}} 
\def\dL{\mathcal{L}_D} 

\def\SLR{SL(2, \mathbb{R})}
\def\SLZ{SL(2, \mathbb{Z})}

\newcommand{\be}{\begin{equation}}
\newcommand{\ee}{\end{equation}}
\newcommand{\ba}{\begin{aligned}}
\newcommand{\ea}{\end{aligned}}

\newcommand{\bea}{\begin{eqnarray}}
\newcommand{\eea}{\end{eqnarray}}

\newcommand{\ch}{\hbox{ch}}
\newcommand{\td}{\hbox{td}}

\def\adj{\mathop{\mathrm{adj}}\nolimits}

\def\Tr{\mathop{\mathrm{Tr}}\nolimits}
\def\tr{\mathop{\mathrm{tr}}\nolimits}

\def\unit{{1\kern-.65ex {\rm l}}}
\def\1{{1\kern-.65ex {\rm l}}}

\def\CY{{\text{CY}}}

\newcount\hour \newcount\minute
\hour=\time \divide \hour by 60
\minute=\time
\def\now{%
\ifnum \hour<13
  \ifnum \hour=0 \advance \hour by 12 \number\hour:\else \number\hour:\fi%
     \ifnum \minute<10 0\fi%
     \number\minute%
\ A.M.%
\else \advance \hour by -12 \number\hour:%
  \ifnum \minute<10 0\fi%
  \number\minute%
  \ P.M.%
\fi%
}

\makeatother

\begin{document}

\baselineskip=18pt  
\numberwithin{equation}{section}  
\allowdisplaybreaks  



\thispagestyle{empty}

\vspace*{0.4cm} 
\begin{center}

 
  {\huge Theories of Class F and Anomalies}

  \vspace*{1.5cm} 
{Craig Lawrie$\,^1$, Dario Martelli$\,^2$, and Sakura Sch\"afer-Nameki$\,^3$}\\

\vspace*{1.0cm} 
{\it $^1$ Institut f\"ur Theoretische Physik, Ruprecht-Karls-Universit\"at,\\
 Philosophenweg 19, 69120 Heidelberg, Germany }\\
  {\tt {gmail:$\,$ craig.lawrie1729}}\\
  
\smallskip

  {\it $^2$ Department of Mathematics, King's College London, \\
      The Strand, London, WC2R 2LS,  UK}\\
      {\tt {kcl.ac.uk}: dario.martelli}\\

\smallskip
{\it $^3$ Mathematical Institute, University of Oxford \\
 Woodstock Road, Oxford, OX2 6GG, United Kingdom}\\
  {\tt {gmail:$\,$sakura.schafer.nameki}}\\

\vspace*{1cm}
\end{center}

\noindent
We consider the 6d $(2,0)$ theory on a fibration by genus $g$ curves, and
dimensionally reduce along the fiber to 4d theories with duality defects. This
generalizes class S theories, for which the fibration is trivial. The
non-trivial fibration in the present setup implies that the gauge couplings of
the 4d theory, which are encoded in the complex structures of the curve, vary
and can undergo S-duality transformations. These monodromies occur around  2d
loci in space-time, the duality defects, above which  the fiber is singular.
The key role that the fibration plays here motivates refering to this setup as
{\it theories of class F}.  In the simplest instance  this gives rise to 4d
$\mathcal{N}=4$ Super-Yang--Mills with space-time dependent coupling that
undergoes $SL(2, \mathbb{Z})$ monodromies.  We determine the anomaly
polynomial for these theories by pushing forward the anomaly polynomial of the
6d $(2,0)$ theory along the fiber. This gives rise to corrections to the
anomaly polynomials of 4d $\mathcal{N}=4$ SYM and theories of class S.  For
the torus case, this analysis is complemented with a field theoretic
derivation of a $U(1)$ anomaly in 4d $\mathcal{N}=4$ SYM.  The corresponding
anomaly polynomial is tested against known expressions of anomalies for
wrapped D3-branes with varying coupling, which are known field theoretically
and from holography. Extensions of the construction to 4d $\mathcal{N}=0$ and
$1$, and 2d theories with varying coupling, are also discussed.

\newpage

\tableofcontents

\section{Introduction}\label{sec:intro}

Theories of class S are 4d $\mathcal{N}=2$ supersymmetric theories, defined by
dimensional reduction with a topological twist of the 6d $(2,0)$
superconformal field theory (SCFT) on a curve $C_{g, n}$ of genus $g$ with $n$
punctures \cite{Gaiotto:2009we}.  In the special case that $g=1$ and $n=0$ the
theory has enhanced supersymmetry and results in 4d maximally supersymmetric
$\mathcal{N}=4$ Super-Yang--Mills (SYM).  The complex structure moduli of the
curve $C_{g,n}$ encode the gauge coupling parameters of the 4d theories. 
{Pants decompositions of $C_{g,n}$ give rise to duality frames of the
class S theories, which are related by the action of the mapping class group, that 
corresponds to the S-duality group of the 4d theory.} For $C_{1,0}= T^2$ the mapping class group is $\SLZ$ and acts on the
complexified coupling $\tau$ of $\mathcal{N}=4$ SYM in the standard way. 

On the other hand it is well-known that 4d $\mathcal{N}=4$ SYM is realized in
terms of the world-volume theory of a stack of D3-branes and thus is embedded
into Type IIB superstring theory. In this context, the $\SLZ$ duality group in
4d is inherited from the self-duality group acting on the axio-dilaton of Type
IIB. In string theory, a generalization of Type IIB to a theory with
space-time dependent axio-dilaton was proposed, and coined F-theory
\cite{Vafa:1996xn,Morrison:1996na,Morrison:1996pp}. The axio-dilaton is
geometrized in terms of the complex structure of an elliptic curve or $T^2$,
which is fibered over the 10d space-time of Type IIB string theory.  The
interesting new physics happens when this elliptic fibration has
singularities, around which the axio-dilaton undergoes monodromies in $\SLZ$.
String-theoretically, these correspond to 7-branes, which can be viewed as a
kind of complex codimension one ``duality defect" in the 10d Type IIB string
theory. 

\begin{figure}
\begin{center}
\includegraphics[width=12.5cm]{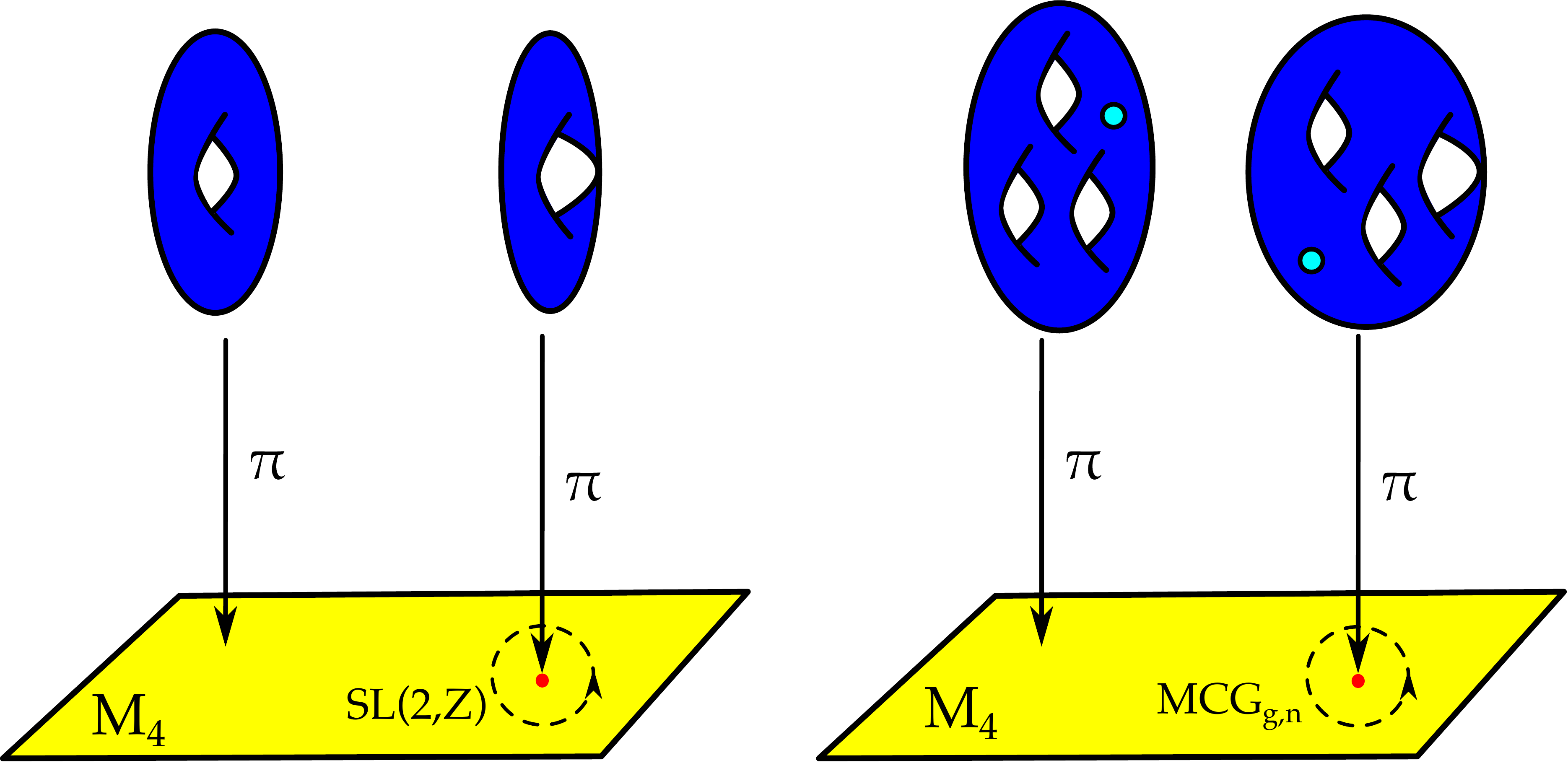}
\caption{Theories of Class F: A sketch of the structure of the 6d space-time for the $(2,0)$ theory, which upon reduction along the fibration $\pi$ gives rise to theories of class F, for $T^2$ and $C_{3,1}$, 
respectively, which are 4d theories on $M_4$ with varying coupling. Singularities in the fiber result in monodromies of the couplings of the 4d theory, which take values in the mapping class group $\text{MCG}_{g,n}$. On the left hand side this is $\SLZ$. In the class F theories the singular loci correspond to duality defects (shown in red), which are real codimension two in $M_4$. 
\label{fig:Fibs}}
\end{center}
\end{figure}

In this paper we will combine the ideas of class S and
F-theory to study a generalization of class S, where the curve $C_{g,n}$ is
non-trivially fibered over the 4d space-time. These are 4d theories where the
coupling is a space-time dependent quantity, and undergoes monodromies in the
duality group. Such theories obtained from 6d $(2,0)$ on a $C_{g,n}$ fibration
by reducing  to 4d along the fiber, will be referred to in short as {\it theories of class F}.  
The simplest class F theories, which already include many interesting features, are obtained 
from $T^2$-fibrations, or equivalently, 4d $\mathcal{N}=4$ SYM with varying coupling and duality defects. 
The second simplest class are $C_{g,0}$-fibrations, which we will also provide a construction of. 
The case including punctures corresponds to fibrations with sections, i.e.~marked points that correspond to maps from the base to the fiber, which carry additional data.
An in depth analysis of these will be deferred to  future work. A useful notation for theories of class F is
\be
T\left[C_{g,n}, \mathcal{F}, G\right]\, :\qquad \mathcal{F} \hbox{ specifies a $C_{g,n}$-fibration over the 4d space-time $M_4$}
\ee
and $G = ADE$ denotes the gauge group of the 6d theory that we start with. In
special cases this reduces to class S, when the fibration is trivial, and
furthermore in this case for $C_{1,0}= T^2$  to 4d $\mathcal{N}=4$. The data
specifying the fibrations will be discussed in depth later on, but it may be
useful to note that for $C_{{1,0}}= T^2$ this data consists of the Weierstrass
line bundle $\mathbb{L}$ and the sections of suitable powers thereof, $f,
g$, that define the Weierstrass model for an elliptic fibration. A sketch of the situation is shown in figure \ref{fig:Fibs}.

In all the cases a construction of the theories in 4d (unless one considers
the abelian theory) will generically result in a non-Lagrangian (or not
obviously Lagrangian) theory, which is intrinsically non-perturbative -- much
like the class S case.  A quantity that characterizes the theories, without
requiring necessarily a perturbative description is the anomaly polynomial
$I_6$.  We will determine $I_6$ for class F theories for both torus and higher genus 
fibrations by pushforward of the $I_8$ anomaly polynomial of the 6d $(2,0)$ theory. 
In the case of $T^2$-fibrations, we can furthermore compare this with a direct anomaly computation 
for a $U(1)$ symmetry that is related to $\SLZ$ and find agreement. The advantage of the pushforward 
description is that it captures intrinsically the contributions from the duality defects as well. This will be 
illustrated for several fibrations.

To substantiate these results we will compare to related configurations, where
e.g.~D3-branes were studied in the context of F-theory compactifications,
either in terms of D3-instantons or wrapped D3-branes, that are strings in the
transverse space-time
\cite{Martucci:2014ema,Haghighat:2015ega,Assel:2016wcr,Lawrie:2016axq,Couzens:2017way,Choi:2017kxf,Couzens:2017nnr,Choi:2018fqw}.
The 2d SCFTs obtained from wrapped D3-branes have 2d $(0,4)$ or $(0,2)$
supersymmetry and are dimensional reductions of class F where the elliptic
fibration $\mathcal{F}$ is non-trivial only over a curve $\Sigma \subset M_4$.
We determine the dimensional reduction of the class F anomaly polynomial and
compare with the known $I_4$ anomaly polynomials of these 2d SCFTs. 

To provide more depth to this proposal,  let us discuss class F in the case of $C= T^2$
in some more detail. This theory has an intrinsically 4d description in terms
of 4d $\mathcal{N} = 4$ SYM, with space-time dependent $\tau$  and with duality
defects of complex codimension one, around which $\tau$ undergoes $\SLZ$ monodromies.  Alternatively, it
is the 6d $(2,0)$ theory on an elliptic fibration with total space $M_6$
\be\label{EllFib}
T^2 \hookrightarrow M_{6} \stackrel{\pi}{\longrightarrow} M_4 \,,
 \ee 
where the complex structure of $T^2$ is $\tau$ and $M_4$ is the 4d space-time (this can be compact or non-compact, Euclidean or Lorentzian).
One defining datum of an elliptic fibration
\cite{MR1078016} is the Weierstrass line bundle  $\mathbb{L}$ on $M_4$, which
is trivial when the fibration is a product, as well as the Weierstrass equation $y^2 = x^3 + fx + g$, determined by $f$ and $g$, which are sections of powers of $\mathbb{L}$ and thus position dependent on $M_4$. In summary, the data defining this theory of class F is 
\be\mathcal{F} = \{\mathbb{L}, f, g\} \,.
\ee
Alternatively we can characterize these class F theories in terms of $\mathcal{N}=4$ SYM, where the complexified gauge coupling 
$\tau$ varies according to the complex structure of the elliptic curve specified by the Weierstrass model, and has duality defects that are characterized in terms of 
\be
\boxed{
\ T\left[T^2, \{\mathbb{L}, f, g\}; G\right]  \quad \equiv \quad  \hbox{4d $\mathcal{N}=4$ SYM, gauge group $G$, $\tau=\tau(f, g)$, duality defects\ }}
\ee
The duality defects are located along 2d 
subspaces of $M_4$, above which the $T^2$-fiber degenerates. The  type of singular $T^2$-fiber furthermore encodes (and in fact classifies) the duality defects (see figure \ref{fig:Fibs}).

The local $U(1)$ symmetry associated with the line bundle has a field theoretic interpretation in 4d $\mathcal{N}=4$, where it is a chiral rotation, that the fermions undergo when applying a duality transformation in $\SLZ$ to the theory \cite{Bachas:1999um, Kapustin:2006pk, Martucci:2014ema, Assel:2016wcr}. The duality acts on the complexified coupling of $\mathcal{N}=4$ SYM by 
\be
\tau \rightarrow {a\tau + b \over c \tau + d} \,,\qquad \gamma = \begin{pmatrix} a & b \\  c & d \end{pmatrix} \in \SLZ \,,
\ee
and needs to be accompanied by a $U(1)_D$ rotation on the fermions (and supercharges)
 \be\label{U1DDef}
U(1)_D: \qquad e^{i \alpha_\gamma (x)} = { |c\tau + d| \over c\tau + d} \,.
\ee
This $U(1)_D$ is gauged and defines a line bundle $\mathcal{L}_D$, with connection that locally takes the form $Q= - {d\tau_1\over 2 \tau_2}$. Chiral fermions in the vector multiplet are charged under this $U(1)_D$ and give rise to a triangle anomaly.
We compute the associated  $c_1(\mathcal{L}_D)$ dependent terms in
the anomaly polynomial $I_6$ of 4d $\mathcal{N}=4$ SYM, and show that these include the anomaly 
\be\label{BBGAno}
\int_{M_4} \Sigma(x) \, \text{tr} R\wedge R \,, 
\ee
where $\Sigma (x)$ is the parameter of the $U(1)_D$ gauge transformation in (\ref{U1DDef}), 
as was noted in \cite{Bachas:1999um}. We show that there are two additional contributions to the anomaly:  one is relevant when there is a non-trivial normal bundle, and depends on the R-symmetry bundle, the other is an explicitly $c_1(\mathcal{L}_D)$-dependent term. 
The conclusion is that the anomaly
polynomial of 4d  $\mathcal{N} = 4$ Super-Yang--Mills on $M_4$ with gauge
group $G$ has {new $c_1(\mathcal{L}_D)$-dependent terms}
\begin{equation}\label{eqn:POLY}
\hbox{{4d $\mathcal{N}=4$ SYM}:}    \qquad  \quad 
   \ba
  I_6  =&\  \frac{1}{2}d_G c_3(\mathcal{S}_{6}^+) -
      \frac{1}{2} r_G c_2(\mathcal{S}_{6}^+) c_1(\mathcal{L}_D) 
      + \frac{1}{4} r_G
           p_1(M_4) c_1(\mathcal{L_{D}}) \cr 
           &  -\frac{61}{4} r_G c_1(\mathcal{L}_D)^3  \,,
           \ea
\end{equation}
where $\mathcal{S}_{6}^+$ is the positive chirality representation of the
$SO(6)$ R-symmetry bundle and 
\be
{c_1(\mathcal{L}_D) = \frac{F_D}{2\pi}}
\ee
is the curvature of
the connection for the $U(1)_D$ bundle. {This expression includes the
contribution to the anomaly from the duality defects, which are most easily
determined by starting in 6d and reducing along the $T^2$ fiber.}
The gauge group data entering the
expression are the rank $r_G$ and dimension $d_G$. These terms become relevant
for the class F theories, which have a non-trivial $F_D$ background, sourced
by a  space-time depending $\tau$. The gauge group data entering the
  anomaly polynomial, the rank and dimension, are invariant under the mapping
  of $G$ to its Langlands dual group $^LG$ {under an S-duality
  transformation}.

Alternatively, following from the definition of class F starting in 6d, we can compute the anomaly by
 pushforward of the $I_8$ anomaly polynomial of the 6d $(2,0)$ theory along
the fiber of the fibration (\ref{EllFib}) and find agreement 
\be
\pi_* I_8 = I_6 \,,
\ee
under the identification of the $U(1)_D$ line bundle $\mathcal{L}_D$ with the Weierstrass bundle $\mathbb{L}$
\begin{equation}
  \mathcal{L}_D = \mathbb{L} \,.
\end{equation}
Once the Weierstrass line $\mathbb{L}$ bundle is fixed, the terms in the first
line of (\ref{eqn:POLY}) are independent of the choice of $f$ and $g$ and thus
in particular of the specific $\tau$-profile. We shall refer to such terms,
that are independent of $f$ and $g$ as {\it $\mathbb{L}$-universal} -- though
often we will abbreviate this to simply {\it universal}.

The $c_1(\mathcal{L}_D)^3$ term,
as we will discuss later on, depends on the number of sections that the
elliptic fibration has\footnote{In (\ref{eqn:POLY}) the coefficient of
$c_1(\mathcal{L}_D)^3$ is that for a fibration with exactly one section.}. Furthermore there can be terms arising from defects
that carry non-abelian flavor symmetries, which are again dependent on the
specific fibration. The reader is referred to section \ref{sec:ClassFNU} for the precise form of these additional terms. We derive these contributions from duality defects  by a careful analysis of the pushforward of $I_8$ for singular elliptic fibrations: the Kodaira singularity type determines  the flavor symmetries $G_F$ of the duality defects \cite{Assel:2016wcr}, and the corresponding terms to $I_6$ in the presence of duality defects that are localized along a surface $S_{G_F}$ are 
\be
I_6^{\text{duality defects}} = {\mathfrak{a}_2 \over 4} \, c_1(\wL)^2 S_{G_F} - {\mathfrak{a}_1 \over 8 }c_1(\wL) \, S_{G_F}^2 +  { \mathfrak{a}_0\over 4} \, S_{G_F}^3  \,,
\ee
where $\mathfrak{a}_i$ depend on the singularity type/flavor symmetry and are
determined in section \ref{sec:ClassFNU}. Here we think of the 6d space-time
as a complex three-fold and $S_{G_F}$ is a divisor (dual to a 2-form). In fact
this implies that duality defects for class F with $T^2$ fibers are classified
in terms of the Kodaira classification of singular elliptic fibers
\cite{Kodaira}.

Class F with higher genus curves can be discussed in a similar way starting in
6d. We explore in this paper the case of no punctures and fibrations that are
realized in terms of projective bundles as plane-curve fibrations. We find a similar generalization of
the anomaly polynomial of class S \cite{Alday:2009qq, Tachikawa:2015bga} to
include correction terms, see (\ref{eqn:classfgap}).  In this case the only
way we can so far analyze this is starting in 6d, and it would indeed be
interesting to complement this with an intrinsically 4d analysis, much like in
the $\mathcal{N}=4$ SYM case. We should remark that these corrections to the
anomaly polynomial are complementary to the ones discussed recently, with
terms depending on forms on the moduli space of the couplings of the theory
\cite{Tachikawa:2017aux,Seiberg:2018ntt}. The terms we obtain become relevant
when the coupling becomes dependent on physical space-time.  As mentioned earlier, punctures
will correspond to sections of the fibrations, which carry additional representation theoretic data. 
For elliptic fibrations, models with small number of sections are well-understood and can be studied systematically. 
For higher genus, already models with one section will correspond to extra data. 
Generalizing the derivation of the anomaly polynomials will be possible by combining the methods of the present paper with those developed for class S with punctures in
\cite{Bah:2018gwc}, which will be discussed elsewhere.
We will in the following abbreviate $C_g \equiv C_{g, 0}$.

The plan of the paper is as follows: {we begin by defining theories of class F in section \ref{sec:ClassF}.}
In section \ref{sec:U1D} we discuss the origin of the $U(1)_D$ symmetry from
6d as well as in 4d $\mathcal{N}=4$ SYM, and its generalization to higher
genus. We then derive the anomaly terms for 4d $\mathcal{N}=4$ SYM that are
$U(1)_D$ dependent from a field theory point of view in section
\ref{sec:from4dspec}. This is complemented in section \ref{sec:M5} by a
pushforward of the 6d anomaly polynomial along the fiber, which we discuss for
both $g=1$ and $g>1$, as well as the contributions from duality defects. We
check our results against known expressions for anomaly polynomials of {2d
theories arising from} D3-branes wrapped on curves in F-theory
compactifications in section \ref{sec:onC}. Finally, some extensions beyond
class F are discussed in section \ref{sec:Extensions}, in particular to the
non-supersymmetric case of 6d self-dual tensor and 4d Maxwell theory, class
S and  $\mathcal{N}=1$ theories in 4d with space-time-dependent coupling. We
conclude in section \ref{sec:CO} and provide several appendices with details
of conventions and computations.


\section{Theories of Class F}
\label{sec:ClassF}

\subsection{F is for Fiber}

Theories of class S are 4d $\mathcal{N}=2$ theories defined as dimensional
reductions with a topological twist of the 6d $(2,0)$ SCFTs with gauge group
$G$ on a Riemann surface $C_{g,n}$, of genus $g$ with $n$ punctures
\cite{Gaiotto:2009we}. The different pair of pants decompositions of the curve 
correspond to different duality frames of the 4d theories. The
simplest precursor of this is the 4d $\mathcal{N}=4$ theory, which is the 6d
$(2,0)$ theory on a $T^2 = \mathbb{E}_\tau$, an elliptic curve, whose  complex
structure is identified as 
\be
\tau = {\theta \over 2 \pi} + i   {4\pi \over g^2} \,.
\ee 
The duality group is $SL(2, \mathbb{Z})$, which acts by modular transformations 
\be
\tau \rightarrow {a\tau + b \over c\tau + d} \,,\qquad  ad-bc =1\,,\quad a,b,c,d\in\mathbb{Z} \,.
\ee
The type of theories we would like to consider generalize this setup to allow for the curve $C$ to vary over space-time -- consistently with the duality symmetries of the theories in 4d.  When $C= T^2$ is an elliptic curve, some aspects of these theories have 
been studied in \cite{Martucci:2014ema, Assel:2016wcr}, where the 4d
space-time was specialized to a compact complex surface, and
\cite{Haghighat:2015ega, Lawrie:2016axq}, where the 4d space-time was
specialized to $\mathbb{R}^{1,1} \times \Sigma$, with $\Sigma$ a complex curve.
For higher genus fibrations a brief discussion has appeared in
\cite{Gadde:2014wma}. 

Our main focus here will be on the 4d theories on $M_4$ and their anomaly polynomials. 
The data that we will use to define the theories are genus-$g$ fibrations over $M_4$
\be
\mathcal{F}: \qquad C_g \ \hookrightarrow \ M_6 \ \rightarrow \ M_4 \,,
\ee
which geometrically model the profile of the complex couplings of the theory of class S on $M_4$. 
For $g>1$ the theory is defined with a topological twist along the fiber.
Subspaces $\Delta \subset M_4$, of real codimension two, above which the
fibers develop singularities correspond to duality defects, around which the couplings
undergo a duality transformation in the mapping class group $\text{MCG}_g$ of
$C_g$. 

For $C=T^2$, the theory in 4d is $\mathcal{N}=4$ SYM with space-time varying complex coupling $\tau$. 
The situation in this case is thus very similar to IIB with varying axio-dilaton, i.e.~F-theory \cite{Vafa:1996xn}, and indeed the theories obtained in this way can be thought of as the world-volume theories of D3-branes in F-theory, where the axio-dilaton variation descends to the variation of $\tau$. The theory can either be studied via M/F-type duality, where the 6d theory is realized in terms of M5-branes, or directly in terms of a 4d theory with varying coupling $\tau$. The latter point of view is advocated in F-theory in \cite{MorrisonTBA}, which defines this theory as Type IIB with a fibration, that is specified by a bundle and sections $f$ and $g$ that specify a Weierstrass model. 

The theories we will study are a generalization of class S to fibrations and
will be referred to as\footnote{Another reason to motivate this name is of
course the similarity to the construction of F-theory, where the axio-dilaton
can be thought of as fibered over space-time as well.} {\it theories of class
F}: they are obtained by reducing the M5-brane theory on a curve $C$, which
however varies over the 4d space-time, allowing for monodromies in duality
symmetries of the 4d theory, which are elements of the mapping class group
$\text{MCG}_g$.
The reduction from 6d to 4d is along the fiber of a fibration $C
\hookrightarrow M_6 \rightarrow M_4$. 
The data that defines these theories is the gauge group $G$ of the 6d theory, together with the fibration. 
Alternatively, in the nomenclature where class S theories are {denoted by} $T[C_g; G]$, class F theories are determined by the data 
\be T\left[C_g, \mathcal{F} ; G\right]\,:\qquad 
\left\{ \ba
C_g\qquad & \text{genus $g$ curve}\cr 
G\qquad & \text{gauge group of ADE type}\cr 
\mathcal{F}\qquad & \text{data specifying a genus $g$-fibration over 4d space-time $M_4$}
\ea
\right. \,.
\ee
As noted earlier, these are generically not Lagrangian theories. 
Nevertheless we will be able to compute the anomaly polynomials of the class F theories for $g=1$ and an infinite subclass of  higher genus theories. One key ingredient that we will leave for the future are punctures, which will be studied elsewhere. 

To fill the description of class F theories with life we will now specify the data required for the fibrations $\mathcal{F}$, and give concrete descriptions of the fibrations in the case $g=1$ and $g>1$, respectively.

\subsection{Elliptic Fibrations}

We start by exploring $C= T^2$ and to emphasize the complex nature will often specify this as an elliptic curve $\mathbb{E}_\tau = \mathbb{C}/\mathbb{Z} \oplus \tau \mathbb{Z}$. The class F theories of this type are  characterized by the data 
\be\label{TEtau}
T\left[\mathbb{E}_\tau,\mathcal{F}; G\right] \,, \qquad  \mathcal{F} = \{\wL, f, g \}\,,
\ee
where $\wL$ is a  line bundle
\be
\wL \rightarrow M_4 \,,
\ee
such that there are sections 
\be
f \in H^0\left(\wL^4\right)\,,\qquad 
g \in H^0 \left(\wL^6\right) \,,
\ee
 and $G$ is the gauge group in 6d.
In this description we assume the fibration to have a section (a map from $M_4$ to the fiber), and thus a description in terms of a Weierstrass equation
\be
y^2 = x^3 + fx + g \,.
\ee
This provides the total space of the fibration $M_6$ with a description in terms of an elliptic fibration over $M_4$.
For trivial fibrations (i.e.~$\wL = \mathcal{O}$), the theory reduces to the standard 4d $\mathcal{N}=4$ SYM theory with gauge group $G$ on $M_4$ with coupling $\tau$ given by the complex structure of the fixed elliptic curve. 

We will begin with $G=U(1)$ where the theory can be understood completely explicitly \cite{Assel:2016wcr} as a dimensional reduction of the tensor multiplet. 
Starting with the 6d $(2,0)$ theory we see that there are two aspects to these 4d theories -- already noted in \cite{Assel:2016wcr}: 
the ``bulk'' part of the spectrum, obtained from the reduction of the tensor multiplet, which results in a 4d $\mathcal{N}=4$ theory with varying coupling. The second contribution arises from singular fibers -- i.e.~they are localized defect modes above which the elliptic fiber becomes singular. In terms of the data in (\ref{TEtau}) this locus is given by the subspaces in $M_4$ that satisfy
\be
\Delta = 4 f^3 + 27 g^2=0\,.
\ee 
It is important to note that the 4d space-time $M_4$ can be compact or non-compact. 

Alternatively we define class F theories with $C= T^2$ directly in 4d, as  4d $\mathcal{N}=4$ SYM with varying coupling, which is consistent with the $SL(2, \mathbb{Z})$ duality group. Again this requires specifying a line bundle $\mathcal{L}_D$ and sections $f$ and $g$, which determine the $\tau$-profile along $M_4$. 
In some cases the coupling will only vary over part of the 4d space-time $M_4$. We will at times indicate the subspace where the coupling varies by superscribing it with $\tau$ 
\be
 M_4 = S^\tau  \quad \hbox{or}  \quad S^\tau \times \mathbb{R}^{1,1} \,,
\ee
where $S^\tau$ is embedded into the base of an elliptic fibration.
Concretely, either $S^\tau= \Sigma$, a complex curve, which will be studied in
section \ref{sec:onC}, or $S^\tau= D$, a complex surface.

\subsection{Fibrations by Genus $g>1$ Curves}\label{sec:Fibg}

Class F which generalizes class S theories, are constructed from fibrations by
genus $g$ curves $C_g$.  There is no canonical realization as in the elliptic
case, however we will consider for simplicity here the case of plane curve
fibrations.  For genus $g$ fibrations, the monodromies around singular fibers
are in the mapping class group $\text{MCG}_g$.  We can in principle include
punctures, as additional data on the curve, {but in this paper we will refrain
  from doing so. 
To be able to define the fibration, $M_4$ will have a complex structure, at least on the subspace of $M_4$ over
which the fibration is non-trivial.
The total space can be realized as a hypersurface in a projective bundle, for
which the following data is required:
\be
\mathcal{F} = \left\{
\ba
\mathcal{J} \rightarrow M_4: &\qquad \text{line bundle over $M_4$} \cr 
\varpi= (a,b,d,e)\in\mathbb{Z}^4 \cr 
\mathbb{P}\mathcal{E} = \mathbb{P}(\mathcal{O} \oplus \mathcal{J}^a \oplus \mathcal{J}^b):& \qquad \text{projective bundle over $M_4$} \cr 
Y_{\varpi} =0 : &\qquad   \text{hypersurface equation in $\mathbb{P}\mathcal{E}$ with} \ [Y_{\varpi}] = d\,  H + e \,
c_1(\mathcal{J})\,.
\ea\right. 
\ee
Here $H$ is the hyperplane class of $\mathbb{P}\mathcal{E}$. This gives a plane-curve fibration where the genus is related to the degree $d$ by
\be
g= {(d-1) (d-2) \over 2}\,.
\label{gendegformula}
\ee
This does not realize all genera, in particular this does not
include the case $g=2$. Genus 2 curves are 
 hyperelliptic and can be realized in terms of a similar construction to the above,
 and a partial list of singular fibers was determined by Ogg and
Namikawa--Ueno \cite{Ogg, NamikawaUeno}. 
More generally, genus $g$ fibrations have a description
as Lefschetz fibrations (over 4d base spaces). This description will be useful
in particular for studying the duality symmetries, however in this paper our
focus is to determine the anomaly polyomials for such theories, which require
us to be able to pushforward forms along the fiber. Such a description is
known to us explicitly only for hypersurfaces in projective bundles, as given above.

Obviously, plane curve fibrations specialize to the elliptic case. Indeed from (\ref{gendegformula}) for $g=1$  and $\varpi= (2,3,3,6)$ this corresponds precisely to the smooth Weierstrass model for an elliptic Calabi-Yau,  with  $\mathcal{J}=\wL$.  
Fibrations with multiple sections can also be realized, e.g.~$\varpi= (1,1,3,3)$ corresponds to  the two-section model.


\section{$U(1)_D$ Symmetry of Class F Theories}
\label{sec:U1D}

{The class F constructions introduced in the last section by elliptic or plane-curve fibrations are each  based on a line bundle. We will now see that the connection on this line bundle corresponds to a $U(1)_D$ connection in the 4d theory, which is sourced by the space-time dependent coupling. In the simplest case of $C=T^2$, writing  $\tau =
\tau_1 + i \tau_2$, this connection is}
\begin{equation}\label{QConn}
  Q = - \frac{1}{2\tau_2} d \tau_1 \,.
\end{equation}
{In this section we will explain the origin of this $U(1)_D$,  from 6d, where
it is related to the local Lorentz symmetry of the fibral curve, or directly in 4d. This $U(1)_D$ will play a key role in that 
it is anomalous and we will show that the anomaly polynomials for 4d $\mathcal{N}=4$ and class S, have corrections depending on the non-trivial curvature of a $U(1)_D$ line bundle. These terms are non-zero whenever we extend these theories to varying coupling, i.e.~to class F. }

\subsection{$U(1)_D$ from 6d $(2,0)$}

We are considering the 6d $(2,0)$ theory on a complex threefold with a
fibration structure, $\pi: M_6 \rightarrow M_4$, where $M_4$ is a (not
necessarily compact)  complex
surface\footnote{{We may take $M_4$ to be a product manifold, where only
the factor over which the fibration is non-trivial is required to be complex.}}. 
{We will assume furthermore, that the threefold is K\"ahler.}
The tangent bundle $TM_6$ to $M_{6}$ has reduced
holonomy $SU(3)_L \times U(1)_L$, likewise the holonomy of $TM_4$ is $SU(2)_l \times U(1)_l$. 

Each of these tangent bundles admits a spin connection,
$\Omega$ and $\omega$ respectively, and
the curvature associated to {the $U(1)$ parts of} these connections are, respectively, $\mathcal{R}_{M_6}$ and
$\mathcal{R}_{M_4}$, such that (see appendix \ref{app:CCs} for our conventions)
\begin{equation}
  c_1(M_n) = \frac{1}{2 \pi} \mathcal{R}_{M_n}    \,.
\end{equation}
When considering a fibration by genus $g$ curves $\pi: M_6 \rightarrow M_4$ there
exists a short exact sequence defining the (rank one) relative tangent bundle
\begin{equation}
  0 \rightarrow T_{M_6/M_4} \rightarrow TM_6 \rightarrow \pi^* TM_4
  \otimes I_X \rightarrow 0
  \,,
\end{equation}
where $I_X$ is a sheaf supported on the singular fibers of the fibration,
and which we shall suppress the details of in the following.
As such we can see that the above curvatures on the two tangent bundles, $TM_6$ and $TM_4$ are related
through 
\begin{equation}
  c_1(M_6) = c_1(M_4) + c_1(T_{M_6/M_4}) + \cdots \,,
\end{equation}
where from now on we will suppress the pullback on the $c_1(M_4)$. In this way
we can see that the $U(1)$ part of the spin connection on $M_{6}$ decomposes as
\begin{equation}
  \Omega = \omega + \mathcal{A}  + \cdots \,,
\end{equation}
where $\mathcal{A}$ is a connection one-form on $T_{M_6/M_4}$. 
For genus one fibrations, we have
\begin{equation}
  T_{M_6/M_4} = \pi^* \wL^\vee \,,
  \label{reltanweis}
\end{equation}
and thus we identify the connection $\mathcal{A}$ with $(-Q)$, the connection on
the dual of the Weierstrass line bundle $\mathbb{L}$. This is the vector potential for the local
$U(1)_D$ abelian symmetry, as was determined explicitly in
\cite{Assel:2016wcr}. It was furthermore shown there that the connection is precisely 
the one in (\ref{QConn}).

For $g > 1$,  $T_{M_6/M_4}$ has a more
  complicated structure, as $TC_g$ is a non-trivial bundle, and the gauge
  connection for the $U(1)_D$ bundle will be mixed, in $\mathcal{A}$, with the 
  potential {$\mathcal{A}_g$} for the {$SO(2)$} holonomy of $C_g$. To see this, consider the case 
of a product space $M_4 \times C_g$,  then
$T_{M_6/M_4}$ is just $TC_g$,  {namely $\mathcal{A} = \mathcal{A}_g$}. 
We do not have a universal formulation for
genus $g$ fibrations, however for the plane curve fibration introduced in section \ref{sec:Fibg} the canonical class of the hypersurface fibration is 
\begin{equation}
  K_{Y_{\bar{w}}} = \pi^*\left( (e - a - b)c_1(\mathcal{J}) - c_1(M_4)\right)
  + (3 - d) c_1(\mathcal{O}_{\mathbb{P}(\mathcal{E})}(1)) \,,
  \label{canclass}
\end{equation}
and the relative tangent bundle
of such fibrations can be seen to contain
a factor of $\mathcal{J}$, \begin{equation}\label{eqn:reltgt}
  T_{M_6/M_4} = \left(\pi^*  \mathcal{J}^{-(e - a - b)} \right) \otimes {\mathcal{R}} 
  \,,
\end{equation}
the connection of which is expected to be the $U(1)_D$ connection, and where this equation defines $\mathcal{R}$.
The pushforward of the relative tangent
bundle is rank $g$, which follows since rank$(T_{M_6/M_4}) = 1$, as the fiber is
a complex curve, and the pushforward of the vector bundle is the rank of the
original bundle with multiplicity given by the degree of the morphism, which
is $g$. 
From (\ref{eqn:reltgt}) and the projection formula it follows that $
\pi_* T_{M_6/M_4} = \mathcal{J}^{-(e - a - b)} \otimes \pi_* (\mathcal{R} )$,
where the pushforward $\pi_* (\mathcal{R} )$ is a rank $g$ bundle. Notice
that for $\varpi=(2,3,3,6)$ the relation (\ref{eqn:reltgt}) reduces precisely
to (\ref{reltanweis}), where the bundle $\mathcal{R}$ is trivial.

\subsection{$U(1)_D$ in 4d}\label{sec:U1d4d}

{In the case of class F theories of type $T[T^2, \mathcal{F}]$ the discussion of the $U(1)_D$ from 6d can be complemented with a direct analysis in 4d.}
The situation is reminiscent to Type IIB with a non-trivial axio-dilaton background, where the {supergravity background} scalars parametrize the coset $SL(2, \mathbb{R})/U(1)$ \cite{Gaberdiel:1998ui, Minasian:2016hoh}. In this case the ungauge-fixed version of the Type IIB theory has three scalars parametrizing $\SLR$ and the chiral fermions transform under a local $U(1)$ symmetry, which upon gauge fixing is identified with $U(1)\subset \SLR$. This gauge fixing removes one of the three scalars, but for this gauge condition to be 
preserved under a general $\SLR$ transformation, these have to be accompanied with a compensating $U(1)$-transformation, which acts non-trivially on the fermions -- in this way an $\SLR$ transformation induces a $U(1)$ action on the fermions. 
This $U(1)_D$ plays a fundamental role in formulating F-theory, i.e.~Type IIB with varying axio-dilaton. 

A very similar situation arises in 4d $\mathcal{N}=4$ SYM, where again there are background fields that parametrize $\SLR /U(1)$, and it is the anomaly of this $U(1)$ symmetry that we will discuss in the following.  
From a 4d point of view the varying $\tau$ background is realized by coupling the theory to 
a non-dynamical off-shell supergravity multiplet \cite{Festuccia:2011ws}, in this case we couple to $\mathcal{N} = 4$ conformal supergravity \cite{Bergshoeff:1980is}. 
This supergravity has an $SL(2,\mathbb{R})$ global and $U(1)$ local symmetry: the scalar manifold has three scalars $\tau_1, \tau_2, \phi$, which parametrize an element of $SL(2, \mathbb{R})$ by 
\be
\Phi =  {1\over \sqrt{\tau_2}}
\begin{pmatrix} 
\tau_2 \cos\phi    + \tau_2 \sin\phi & - \tau_2\sin\phi + \tau_1 \cos\phi\\
\sin\phi & \cos\phi  
\end{pmatrix}  \in SL(2,\mathbb{R}) \,.
\ee
The symmetries $h\in SL(2,\mathbb{R})$ and $R(\alpha_\gamma(x)) \in U(1)$ act by 
\be
\Phi \rightarrow  h \Phi \begin{pmatrix} \cos \alpha_\gamma  & \sin
\alpha_\gamma \\  -\sin \alpha_\gamma & \cos \alpha_\gamma\end{pmatrix}\,.
\ee
A useful gauge fixing of the $U(1)$ was described in \cite{Bachas:1999um} in analogy to the one in IIB \cite{Gaberdiel:1998ui}, by fixing $\phi=0$ (under the local  $U(1)$ this shifts by $\phi \rightarrow \phi + \Sigma$), and  identifies $\Phi$ with the complexified coupling $\tau$, which takes values in the coset $\SLR/U(1)$.
This gauge choice is invariant under $SL(2, \mathbb{R})\times U(1)$ only with the additional compensating (local) $U(1)$ transformation \be
U(1)_D:\qquad e^{i \alpha_\gamma (x)} = { |c\tau + d| \over c\tau + d} \,,
\ee
where $h \tau = {a\tau + b \over c\tau + d}$, and this acts by shift on $\phi$, and a phase rotation on the fermions. 
This is precisely the $U(1)_D$ symmetry that also descends from 6d \cite{Assel:2016wcr}, and was observed to be relevant for S-duality transformations of 4d $\mathcal{N}=4$ SYM in \cite{Kapustin:2006pk}.
The gauge potential for the local $U(1)_D$ symmetry is
\be\label{QDef}
\partial_\mu \phi + Q_\mu =\partial_\mu \phi - {\partial_\mu \tau_1 \over 2\tau_2}  \,,
\ee
which is precisely the one obtained earlier from 6d in (\ref{QConn}).
There is triangle anomaly for the $U(1)_D$ current, 
involving the gauginos in the loop which transform under the $U(1)$
symmetry. One such contribution was computed in \cite{Bachas:1999um}, in the
case of trivial normal bundle, where it was shown that  the D3-brane on $M_4$ has an anomaly under the $U(1)$
symmetry, arising from the coupling to gravitons, of the form
\begin{equation}
  \delta S_{\text{gauginos}} \propto \int_{M_4} \Sigma(x)  \, p_1(M_4) \,,
\end{equation}
where as before $\Sigma$ is the gauge variation of $\phi$ under the local 
$U(1)$ transformation.
Such an anomalous variation is cancelled by the addition of a counterterm to
the 4d action
\begin{equation}\label{BBGct}
  S_{ct} = \int_{M_4} \phi\,  p_1(M_4) \,.
\end{equation}
In this paper we extend this result to include a non-trivial R-symmetry
bundle, and to include $U(1)^3$ anomalies, and in addition we study the
contribution to the $U(1)$ anomalies, not just of the bulk $\mathcal{N} = 4$
gauginos, but also those from the degrees of freedom living on the duality
defects when $\tau$ varies holomorphically along $M_4$.

The classical $\SLR$ is broken to $\SLZ$ quantum mechanically, and in fact the full duality group has recently been discussed to be the $\mathbb{Z}_2$ central extension of $\SLZ$, the metaplectic group  \cite{Pantev:2016nze}, see also \cite{MorrisonTBA}, which includes precisely also the action on the fermions with the phase rotation with half-integral charge.


\section{4d $\mathcal{N}=4$ and the $U(1)_D$ Anomaly}
\label{sec:from4dspec}

This section will be entirely about class F theories with $C= T^2$. We begin
by discussing the anomaly in $\mathcal{N}=4$ SYM in 4d associated to the
$U(1)_D$ symmetry. This is a priori independent of the class F
construction from 6d, but what we will see is that there are additional terms
in the anomaly polynomial for $\mathcal{N}=4$ SYM {that} become relevant
whenever there is a non-trivial $U(1)_D$ connection, as in the case of class F. 

We first review the action of the $U(1)_D$ introduced in section \ref{sec:U1D}
on the chiral fermions in the $\mathcal{N} = 4$ vector multiplet, and 
then compute the one-loop anomaly arising from the chiral fermions
on the generic point of the Coulomb branch of any $\mathcal{N} = 4$ SYM
theory, and use 't Hooft anomaly matching to relate to the anomaly at the
origin of the Coulomb branch. This involves a conjecture that any interaction
terms in the effective theory when integrating out the massive fermions and
moving onto the Coulomb branch are irrelevant for any $U(1)_D$-related
anomaly. This is because we expect that the interaction terms will provide a
contribution to the anomaly polynomial proportional to the number of massive W-bosons on
the Coulomb branch; in section \ref{sec:M5} we will see that any
contributions to the $U(1)_D$ anomalies that scale in such a way will be at
odds with the
point of view from the 6d $(2,0)$ superconformal field theory.

\subsection{Anomaly Polynomial for 4d $\mathcal{N}=4$ SYM with Gauged $U(1)_D$}

The $U(1)_D$ symmetry of $\mathcal{N} = 4$ SYM discussed in section
\ref{sec:U1D}, arising from the $SL(2,\mathbb{R}) \times U(1)$ enhanced
symmetry when coupling the 4d theory to an arbitrary supergravity background,
can be seen to act as an R-symmetry of the superconformal group $PSU(2,2|4)$.
In \cite{Intriligator:1998ig,Intriligator:1999ff}\footnote{While in this paper
and in the recent literature this group has been referred to as $U(1)_D$, the
notation in \cite{Intriligator:1998ig,Intriligator:1999ff} is $U(1)_Y$.} it
was shown that $PSU(2,2|4)$ admits such an outer automorphism.
For the abelian 4d
$\mathcal{N}=4$ SYM theory, this gives rise to a symmetry of the spectrum of {some}
observables, though not of the Lagrangian \cite{Intriligator:1998ig}.  The gauginos transform under the
bosonic subgroup $Spin(1,3) \times SU(4)_R$ of $PSU(2,2|4)$ and the duality
$U(1)_D$ as follows
\begin{equation}\label{eqn:refN4ferm}
  ({\bf 2,1,4})_{1/2} \oplus ({\bf 1,2,\overline{4}})_{-1/2}  \,.
\end{equation}
As such, we can compute the anomalies {of the}
$U(1)_D$ symmetry in the usual way, specifically we use the
Wess--Zumino descent procedure (for reviews on this topic see e.g.~\cite{Bilal:2008qx}), where the anomaly due to a chiral
fermion transforming in a representation ${\bf R}$ of a group $G$ is
determined via the anomaly (4+2)-form
\begin{equation}
\left.  \ch(F_{\bf R}) \widehat{A}(TM_4)\right|_{6\text{-form}} \,,
\end{equation}
where $TM_4$ is the tangent bundle to the 4d space-time, $M_4$. 
Applied to 4d $\mathcal{N} = 4$ SYM via the fermions in (\ref{eqn:refN4ferm}) this results in 
\begin{equation}\label{eqn:anom1}
  \frac{1}{2}\left( \ch(\mathcal{S}_6^+ \otimes \mathcal{L}_D^{1/2}) - \ch(\mathcal{S}_6^-
  \otimes \mathcal{L}_D^{-1/2})\right)\widehat{A}(TM_4) = \ch(\mathcal{S}_6^+
  \otimes \mathcal{L}_D^{1/2}) \widehat{A}(TM_4) \,,
\end{equation}
where $\mathcal{S}_6^\pm$ are the complex conjugate vector bundles associated to the
${\bf 4}$, ${\bf \overline{4}}$ of $SU(4)_R$, and $\mathcal{L}_D$ is the
bundle associated to the charge $+1$ representation of $U(1)_D$.
$\mathcal{L}_D$ is the bundle with connection (\ref{QConn}).
This connection e.g.~can be derived by considering a supergravity background for the 4d $\mathcal{N}=4$ theory
with a non-trivial $\tau$ profile \cite{Martucci:2014ema, Maxfield:2016lok, Couzens:2017way}.
If the coupling does not vary over space-time, the 
bundle $\mathcal{L}_D$ is trivial and we recover the expected result for the
anomaly polynomial for constant coupling $\mathcal{N} = 4$ SYM.

We can now compute the anomaly polynomials, starting with the abelian theory $G=U(1)$.
Expanding out the characteristic classes in (\ref{eqn:anom1}), for a summary of our conventions see appendix \ref{app:CCs}, 
one finds that the contribution to the anomaly from the gauginos in the $U(1)$ $\mathcal{N} = 4$ vector multiplet is
\begin{equation}
  I_6^F = \frac{1}{2} c_3(\mathcal{S}_6^+) - \frac{1}{2} c_2(\mathcal{S}_6^+) c_1(\mathcal{L}_D) +
  \frac{1}{12} c_1(\mathcal{L}_D)^3 - \frac{1}{12} p_1(TM_4)c_1(\mathcal{L}_D)
  \,.
  \label{abeliananom}
\end{equation}
The first term is the standard contribution due to the R-symmetry $SU(4)_R$ anomaly. The remaining terms are explicitly $\mathcal{L}_D$ dependent and signal anomalies due to gauging $U(1)_D$.

To generalize this to a $U(N)$ gauge group we first consider the theory in a Higgsed phase. 
On the Coulomb branch of the $U(N)$ theory the gauge group is $U(1)^N$ and the
total number of massless Weyl fermions is {$4N$, where} $N$ is the rank of the
Cartan subalgebra. As such the anomaly contribution is simply\footnote{{The
    expression (\ref{abeliananom}) includes already the factor of $4$ arising
    from the four fermions in the ${\cal N}=4$ vector multiplet. Equivalently, in ${\cal N}=1$ notation, there is
one chiral fermion in a vector multiplet and one in each of three
adjoint chiral multiplets.}} $N \times I_6^F$.  However, this is not the whole
story, as described in \cite{Intriligator:2000eq}, there are Wess--Zumino
interaction terms\footnote{In a similar manner, Green--Schwarz interaction
  terms have recently been utilised in
  \cite{Ohmori:2014kda,Intriligator:2014eaa} to study the anomaly polynomial
  from the Coulomb branch of 6d $\mathcal{N} = (1,0)$ SCFTs.} that are induced
  in integrating out the massive states when moving onto the Coulomb branch,
  and these will modify the anomaly polynomial even deep on the Coulomb
  branch. We shall assume that, also in the case of a space-time dependent
  $\tau$, these Wess--Zumino terms introduce the same contribution to the
  anomaly polynomial as in the constant $\tau$ case, to wit, 
\begin{equation}
  I_6^{WZ} = \frac{1}{2} (N^2 - N) c_3(\mathcal{S}_6^+) \,.
\end{equation}
As such, the anomaly
polynomial of $U(N)$ $\mathcal{N} = 4$ SYM {arising from the bulk degrees
of freedom} is conjectured to be
\begin{equation}\label{eqn:anomaly6das}
  \begin{aligned}
  I_6 &= N I_6^F + I_6^{WZ} \cr
  &= \frac{1}{2} N^2 c_3(\mathcal{S}_6^+) - \frac{1}{2}
  N c_2(\mathcal{S}_6^+)
      c_1(\mathcal{L}_D) + \frac{1}{12} N c_1(\mathcal{L}_D)^3 - \frac{1}{12} N
         p_1({TM_4}) c_1(\mathcal{L}_D) \,.
  \end{aligned}
\end{equation}
As described in section \ref{sec:ClassF}, there are in addition to
  the bulk modes, localised defect degrees of freedom in the theory with
  non-trivial $\mathcal{L}_D$, and these will further contribute to the 
anomaly polynomial.
We will compare (\ref{eqn:anomaly6das}) in the next section to the 't Hooft anomaly coefficients. 

\subsection{'t Hooft Anomaly Coefficients}

Alternatively we can compute the 't Hooft anomaly coefficients directly on the
Coulomb branch of the $G = U(N)$ $\mathcal{N}=4$ SYM
\cite{Intriligator:2000eq}, which consists of $N$ massless $\mathcal{N} = 4$
$U(1)$ vector multiplets, each of which can be written in $\mathcal{N} = 1$
language as a vector multiplet, $V$, and three chiral multiplets, $\Phi_i$.
The R-charges of these multiplets are
\begin{equation}\label{eqn:Rcharges}
  R[V] = 1 \,, \quad R[\Phi_i] = \frac{2}{3} \,.
\end{equation}
{The R-charge of three adjoint scalars $\Phi_i$ is fixed by requiring the cubic superpotential to have R-charge 2, while the R-charge of the vector multiplet 
is such that the gauge field is uncharged}\footnote{{Thus $1$ is the R-charge of the gaugino, while the fermions in the chiral multiplets have R-charge $-1/3$.}}. Equivalently,  these  charges can be derived by decomposing the $SU(4)$ R-symmetry as follows. 
Let us consider the fermions in the representation
$({\bf 2,1,4})$
 under $SO(1,3) \times SU(4)_R$. First we consider
the decomposition to the R-symmetry of an $\mathcal{N} = 2$ subalgebra 
\begin{equation}
  \begin{aligned}
    SU(4) &\rightarrow SU(2)_R \times SU(2)_L \times U(1)_R \cr
    {\bf 4} &\rightarrow ({\bf 2,1})_1 \oplus ({\bf 1,2})_{-1} \,,
  \end{aligned}
\end{equation}
where $SU(2)_R \times U(1)_R$ now forms the R-symmetry of the $\mathcal{N} =
2$ SCFT. We can now apply the relation between the $\mathcal{N} = 2$
R-symmetry and the $\mathcal{N} = 1$ R-charge,
\begin{equation}
  R_{\mathcal{N}=1} = \frac{1}{3}R_{\mathcal{N}=2} + \frac{4}{3} L_3 \,,
\end{equation} 
where $L_a$ are the generators of the $SU(2)_R$ and $L_3$ is such that the ${\bf
2}$ has charges $\pm \frac{1}{2}$, to find
\begin{equation}
  ({\bf 2,1,4}) \rightarrow ({\bf 2,1})_{1} \oplus ({\bf 2,1})_{-1/3} \oplus
  ({\bf 1,2})_{-1/3} \oplus ({\bf 1,2})_{-1/3} \,,
\end{equation}
where the final subscript is the $U(1)_R^{\mathcal{N}=1}$ charge. Since the
R-charge of the fermions in the chiral multiplets is one less than the charge
of the scalar we can see that it follows that (\ref{eqn:Rcharges}) are the
R-charges.  
 Further, since we are considering the fermions in the positive chirality
$({\bf 2,1,4})$ representation we must take the $U(1)_D$ charge of all the fermions to be 
\begin{equation}
  q_D = \frac{1}{2} \,.
\end{equation}
The anomaly polynomial of the ${\cal N}=4$ theory  takes the form
\begin{equation}
\begin{aligned}
 I_{6} =  & \frac{1}{6}    k_{RRR} c_1(R)^3 + \frac{1}{2}k_{RRD} c_{1}(R)^2  c_{1}({\cal L}_D)  +{\frac{1}{6}}k_{DDD}c_{1}({\cal L}_D)^3- \frac{1}{24} k_{D} c_{1}({\cal L}_D) p_{1}({TM_4})\, ,
\end{aligned}
\end{equation}
where recall that the 't Hooft anomaly coefficients are defined as
\begin{equation}
  k_{IJK} = \Tr_{f_+} q_I q_J q_K \,, \qquad k_I =
  \Tr_{f_+} q_I \,,
\end{equation}
where the trace is taken over all positive chirality Weyl fermions, 
and we have written only the non-zero terms.
In particular, the following quantities may be computed directly from the
field content
\begin{equation}\label{eqn:tHooftonCB}
  \begin{aligned}
    k_R &= N \left(3 \times \left(-\frac{1}{3}\right) + 1 \times 1\right) = 0 \cr
    k_{RRR} &= N \left(3 \times \left(- \frac{1}{3}\right)^3 + 1 \times 1^3\right) = \frac{8}{9} N
    \cr
    k_D &= N \left( 4 \times \left(\frac{1}{2}\right)\right) = 2 N \cr
    k_{RRD} &= N \left( 3 \times
    \left(-\frac{1}{3}\right)^2\left(\frac{1}{2}\right) + 1 \times 1^2
    \left(\frac{1}{2}\right)\right) = \frac{2}{3} N  \cr
    k_{RDD} &= N \left( 3 \times
    \left(-\frac{1}{3}\right)\left(\frac{1}{2}\right)^2 + 1 \times 1
    \left(\frac{1}{2}\right)^2\right) = 0  \cr
    k_{DDD} &= N \left( 4 \times \left(\frac{1}{2}\right)^3 \right) =
    \frac{1}{2} N \,.
  \end{aligned}
\end{equation}
As explained previously, Wess--Zumino terms are introduced via the integrating
out of the massive modes as one moves onto the Coulomb branch, these
couplings contribute to the $k_{RRR}$ 't Hooft anomaly coefficient like
\begin{equation}
  n_V \times (\text{contribution from single fermion in ${\bf 4}$}) = \frac{8}{9} (N^2
  - N) \,,
\end{equation}
where $n_V$ is the number of massive fermions. Thus the total $k_{RRR}$ anomaly
coefficient can be calculated on the Coulomb branch to be
\begin{equation}
  k_{RRR} = \frac{8}{9} N + \frac{8}{9} (N^2 - N) = \frac{8}{9} N^2 \,.
\end{equation}

For $\mathcal{N} = 4$ super-Yang--Mills we can use the relationship between
the central charges and the 't Hooft anomaly coefficients
\begin{equation}\label{eqn:ck}
    a = \frac{9}{32} k_{RRR} - \frac{3}{32} k_R \,, \quad c = \frac{9}{32}
    k_{RRR} - \frac{5}{32} k_R \,,
\end{equation}
to determine the central charges for $G = U(N)$ to be
\begin{equation}
  a = c = \frac{1}{4} N^2 \,.
\end{equation}
{It can be verified that, upon  rewriting the $SU(4)_R$ in term of  the $U(1)_R^{\mathcal{N}=1}$, the 't Hooft anomalies computed from the field content in
(\ref{eqn:tHooftonCB}) match exactly with those in (\ref{eqn:anomaly6das}).}

\subsection{Modular Anomaly Revisited}

The Montonen--Olive duality group of 4d $\mathcal{N} = 4$ SYM is
$SL(2,\mathbb{Z})$ \cite{Montonen:1977sn,Osborn:1979tq,Witten:1979ey}, and
  under the S-duality transformations of this group the partition function,
  $Z(\tau)$, is known to transform as a modular form
  \cite{Vafa:1994tf,Witten:1995gf}. This failure of invariance of the
  partition function under the action of the duality group is known as the
  modular anomaly.

When the $\mathcal{N} = 4$ SYM is embedded into string theory, as the
worldvolume theory on a stack of D3 branes, it is expected that the modular
anomaly, which is then the anomaly of a local $SL(2,\mathbb{Z})$, is
cancelled. The putative $SL(2,\mathbb{Z})$ duality group
\footnote{In fact, the
  duality group is an extension of this by including the non-perturbative symmetry  $(-1)^F$ (which extends it to the
  metaplectic group, $Mp(2,\mathbb{Z})$ 
  \cite{Pantev:2016nze}) and the perturbative $\Omega$ and $(-1)^{F_L}$ symmetries, which together extend $SL(2,\mathbb{Z})$ to the Pin$^+$-version of  the double-cover of $GL(2,\mathbb{Z})$ \cite{Tachikawa:2018njr}.}. This subtletly will not concern
us in this discussion.} of Type IIB string theory, from which descends the
$SL(2,\mathbb{Z})$ on the D3-brane, arises as the remnant, after quantisation,
of the $SL(2,\mathbb{R})$ global symmetry of Type IIB supergravity. In
\cite{Bachas:1999um} it was shown, in a particular simple background, that 
part of the modular anomaly is cancelled by considering the D-instanton
corrections to the D3-brane action, together with the counterterm cancelling
the local $U(1)$ anomaly discussed in section \ref{sec:U1d4d}. In this case
the remaining part of the modular anomaly is just proportional to a constant
multiple of $p_1(M_4)$, which is then required to be an appropriate factor of
$2\pi$ for the theory to be consistent under the $SL(2,\mathbb{Z})$.
  
In particular we have extended the discussion in \cite{Bachas:1999um} by
considering a non-trivial normal bundle, implying that there is a triangle
anomaly involving the local $U(1)$ currents and the R-symmetry currents, and
by including the contribution to the overall anomaly from three $U(1)$
currents.

It is necessary that the local $U(1)$ symmetry be non-anomalous, in order to
gauge fix and combine any $SL(2,\mathbb{R})$ transformation with a
compensating $U(1)$ transformation that fixes the gauge. As we saw in
(\ref{eqn:POLY}), the anomaly polynomial, $I_6$, contains terms proportional
to the field strength of the $U(1)_D$ gauge field,
\begin{equation}
 { \frac{F_D}{2\pi} = c_1(\dL) }\,.
\end{equation}
Locally we can always write $F_D$ as 
\begin{equation}
  F_D = d Q \,,
\end{equation}
where $Q$ is defined in (\ref{QDef}).
{The anomaly  associated to a $U(1)_D$ gauge variation with parameter
$\Sigma(x)$ is determined by the descent procedure from $I_6$, and comprises the following terms}
\begin{equation}
  \begin{aligned}
    \int_{M_4} \left( {\frac{F_D}{2\pi}} \wedge p_1(M_4) \right)^{(1)} &= \int_{M_4}
    \Sigma(x) p_1(M_4)\, , \cr
    \int_{M_4} \left(  {\frac{F_D}{2\pi}} \wedge c_2(\mathcal{S}_6^+) \right)^{(1)} &= \int_{M_4}
    \Sigma(x) c_2(\mathcal{S}_6^+) \, ,\cr
    \int_{M_4} \left(  {\left(\frac{F_D}{2\pi}\right)^3} \right)^{(1)} &= \int_{M_4}
    \Sigma(x)  {\left(\frac{F_D}{2\pi}\right)^2} \,.
  \end{aligned}
\end{equation}
This anomaly can be cancelled by adding a local counterterm of the form 
\begin{equation}\label{Fullct}
S_{\text{ct}} = N \int_{M_4} \phi \left({-{1\over 12}} p_1(M_4) -{1\over 2} c_2(\mathcal{S}_6^+) + {1\over 12}\left(\frac{F_D}{2\pi}\right)^2 \right) \,,
\end{equation}
where $\phi$ is as in section \ref{sec:U1d4d}. 
The first term in the above is {exactly} the one that was obtained in
\cite{Bachas:1999um}. There are  two additional terms that we have obtained: one related to the
R-symmetry, the other to the $U(1)_D$ connection, which will be relevant for
non-trivial normal bundles and space-time dependent couplings, respectively. 
The counterterms are not manifestly $\SLZ$ invariant. In \cite{Bachas:1999um}
it was conjectured that there are infinitely many D-instanton corrections,
which result in a pre-factor in (\ref{BBGct}) that makes the term manifestly
modular invariant. The counterterm (\ref{Fullct}) yields a modular anomaly, because under an S transformation of $\phi$, it transforms as
\be\label{eqn:noddy}
N \int_{M_4} \log \left({\tau\over \bar\tau} \right){\left({-{1\over 12}} p_1(M_4) -{1\over 2} c_2(\mathcal{S}_6^+) + {1\over 12}F_D^2 \right)} \,.
\ee

{We note that there is an uncanny resemblance 
between the counterterm (\ref{Fullct}) and the $U(1)$ anomaly counterterm obtained for the 10d Type IIB string in \cite{Gaberdiel:1998ui, Minasian:2016hoh}. In particular,
the counterterm presented in the latter references reads
\be\label{Raffaello}
S_{\text{ct}}^{\text{IIB}} \supset \int \phi \, \left( {1\over 48} p_1(M_{10}) - {1\over 32} \mathfrak{F}^2 \right) \mathfrak{F}^3  \,,
\ee
where $\mathfrak{F}$ plays the same role as $\frac{F_D}{2\pi}$  in the class F theories, but with $\tau$ identified with the Type IIB axio-dilaton.
Although  the  anomalies are not obviously related, it is tempting to speculate that (\ref{Fullct}) and (\ref{Raffaello}) may be related by an 
inflow mechanism analogous to that involving 
M5 branes in M-theory \cite{Freed:1998tg}. 
We also note that, similarly to  (\ref{Fullct}), also the counterterm derived in \cite{Gaberdiel:1998ui, Minasian:2016hoh} does not capture the contribution of defect modes, namely of 7-branes. In the next section we will show how in our context these contributions can be calculated in detail, starting from a 6d theory.}

\section{Anomaly Polynomial of Class F  from 6d}
\label{sec:M5}

In this section we derive the anomaly polynomials for class F for $C = T^2$ and $C_g$, respectively, 
starting with the 6d $(2,0)$ theory and reducing along the fiberal curve $C$. In the case of $C= T^2$ this gives
complete agreement with the 4d field theory analysis of the previous section.

\subsection{Class F with Torus-Fibers: $\mathbb{L}$-Universal Contributions}\label{sec:M5T2}

We begin with class F and $C= T^2$ and consider the contibutions to the anomaly in 4d 
without specifying the precise space-time geometry. This is done by integrating
the $I_8$ polynomial of the 6d theory ``along the fiber" -- more precisely, we
will pushforward the $I_8$ eight-form to a six-form on the base of the elliptic
fibration, that being 4d space-time. This six-form on $M_4$ will then be a
part\footnote{As there may be emergent symmetries in the limit where the
  volume of the fiber shrinks to zero this six-form my not be the full anomaly
  polynomial of the 4d theory, as it may not be sensitive to the anomalies of
these emergent global symmetries.}
of the anomaly polynomial of the 4d $\mathcal{N} = 4$ SYM with varying
coupling on $M_4$.

The anomaly polynomial of the 6d $(2,0)$ theory of type $G$ is
\cite{
Harvey:1998bx
,Intriligator:2000eq,Yi:2001bz,Ohmori:2014kda
}
\begin{equation}\label{eqn:I8ap}
  I_8 = \frac{r_G}{48}\left[ p_2(N_5) - p_2(TM_6) + \frac{1}{4}\left(p_1(N_5)
  - p_1(TM_6)\right)^2 \right] + \frac{h_G^\vee d_G}{24} p_2(N_5) \,.
\end{equation}
Here $N_5$ is the $SO(5)$ R-symmetry bundle {of which the scalars in the tensor
multiplet transform as sections}, $TM_6$ is the tangent bundle to the
six-dimensional worldvolume of the $(2,0)$ theory, and $r_G$, $d_G$, and
$h_G^\vee$ are, respectively, the rank, dimension, and dual Coxeter number of
the $ADE$ gauge group $G$. 

When we consider a compactification of the 6d theory on a $T^2$ the
R-symmetry group in fact enhances, rather than reduces as for the generic
$C_g$ compactification, as the additional scalar from the
compactification combines with the ${\bf 5}$ scalars from the $(2,0)$ theory to give
an $SO(6)_R$ R-symmetry group. This $SO(6)_R$ R-symmetry group is emergent in the
low energy theory, and thus we would not expect to see the full
anomaly associated to this global symmetry. The $SO(5)_R$ R-symmetry of the
6d  theory {is} a subgroup of $SO(6)_R$ corresponding to the branching
rule 
\begin{equation}
  \begin{aligned}
    SO(6)_R &\  \rightarrow\  SO(5)_R \cr
    {\bf 4}, {\bf \overline{4}} &\  \mapsto \ {\bf 4} \,.
  \end{aligned}
\end{equation}
For the $SO(5)_R$ symmetry to be a subgroup of the $SO(6)_R$ symmetry then it
is necessary that the bundle $N_5$, defined on $M_6$, is in fact the pullback
of a bundle, $N_5^\prime$, defined on the base $M_4$
\begin{equation}
  N_5 = \pi^* N_5' \,.
\end{equation}
If we write $\mathcal{S}(N_5^\prime)$ to denote the $SO(5)_R$ spin bundle, and, as in section
\ref{sec:from4dspec}, $\mathcal{S}_6^\pm$ to denote the $SO(6)_R$ spin bundles then we {have}
that 
\begin{equation}
  c_2(\mathcal{S}(N_5^\prime)) = c_2\left(\mathcal{S}_6^\pm\right) \,.
\end{equation}
We will be able to recognise any contribution to the 4d anomaly
six-form that is proportional to $c_2\left(\mathcal{S}_6^\pm\right)$, but not to 
$c_3\left(\mathcal{S}_6^\pm\right)$, which is not visible from the point of view of the
$SO(5)$ subbundle.

We are considering an elliptic fibration $\pi: M_6 \rightarrow M_4$ on which $I_8$
is defined. We can consider the integral over the fiber of this eight-form by
pushing-forward $I_8$ to a six-form on the base, $M_4$. {In particular},
we must compute  $\pi_* I_8$ {where}
\begin{equation}
  \pi_*I_8 = \frac{r_G}{48} \left[ \pi_* \left(-p_2(M_6) + \frac{1}{4}p_1(M_6)^2
  \right) - \frac{1}{2}p_1(N_5) \pi_*p_1(M_6) \right] \,.
\end{equation}
To determine the above two pushforwards we first introduce an auxiliary
complex curve $Z$ and rewrite the Pontryagin classes in terms of Pontryagin
classes of $M_6 \times Z$ and $Z$. We then pushforward these eight-forms on
the product elliptic fourfold, $M_6 \times Z$, using the formulae in
\cite{MR3094035} (see also \cite{Aluffi:2007sx, Aluffi:2009tm, Esole:2011cn}).
For an elliptic fourfold it is
shown in appendix \ref{app:CCs} that 
\begin{equation}
  \pi_*(1 + c_1(Y) + c_2(Y) + c_3(Y) + c_4(Y)) = 12 c_1(\wL)c(B)(1 +
  \mathcal{O}(c_1(\wL),\cdots)) \,,
\end{equation}
where $\mathcal{O}(c_1(\wL),\cdots)$ represents terms that are at least linear
in $c_1(\wL)$ or additional divisors related to codimension one
singular fibers of the fibration. In such manner one can determine
\begin{equation}
  \begin{aligned}
    \pi_* \left( p_1(M_6) \right) &= -24 c_1(\wL) \, ,\cr
    \pi_* \left( -p_2(M_6) + \frac{1}{4}p_1(M_6)^2 \right) &= 12 c_1(\wL)
    p_1(M_4) + \cdots \,,
  \end{aligned}
\end{equation}
and 
thus we conclude that 
\begin{equation}\label{eqn:benandholly}
  \begin{aligned}
    I_6 = \pi_*I_8 &= \left[ \frac{1}{4} r_G p_1(N_5^\prime) c_1(\wL) +
    \frac{1}{4}r_G c_1(\wL)p_1(B) + \cdots \right] \cr
    &= \left[ -\frac{1}{2} r_G c_2(\mathcal{S}(N_5^\prime)) c_1(\wL) -
    \frac{1}{24}(-6 r_G) c_1(\wL)p_1(B) + \cdots \right]
          \,,
  \end{aligned}
\end{equation}
where in the last line we have used that
\begin{equation}\label{eqn:rosieandjim}
  p_1(N_5^\prime) = 2 \ch_2(\mathcal{S}(N_5^\prime)) \,.
\end{equation}

We must now compare this $I_6$ to the one determined from the bulk spectrum of  $\mathcal{N} =
4$ with varying coupling  in (\ref{eqn:anomaly6das}). As we have
already stated, we do not expect to see the $c_3(\mathcal{S}_6^+)$ term in
(\ref{eqn:anomaly6das}) from the integrated 6d anomaly polynomial, as it is
not visible through the $SO(5)_R$ subbundle of $SO(6)_R$. The first thing
  to note is that the anomalies related to the $c_1(\mathbb{L})$, the $U(1)_D$
  anomalies, are proportional to the rank of the 4d gauge group, $r_G$. This
  hearkens back to the statement at the opening of section
  \ref{sec:from4dspec}, where we assumed that, from a 4d point of view, there
  would be no contributions from Wess--Zumino interactions terms. We expected
  that if these terms did contribute then the 't Hooft anomalies would scale
  like $d_G$, which is not what is observed from the anomaly from the 6d
  theory in (\ref{eqn:benandholly}).

The first term in (\ref{eqn:benandholly}) appears identically in
(\ref{eqn:anomaly6das}), however, the second term, does not match. 
The difference is 
\begin{equation}
  I_6^{\text{defects}} = - \frac{1}{24} (- 8 r_G ) c_1(\mathcal{L}_D) p_1(M_4) \,,
\end{equation}
which is not
unexpected -- the 6d theory is sensitive to the defects in
the spectrum, which modify the mixed $U(1)_D$-gravitational anomaly and are not taken into
account by the bulk spectrum computation which leads to (\ref{eqn:anomaly6das}) in
section \ref{sec:from4dspec}. {This is the $\mathbb{L}$-universal contribution of the
duality defects to the $U(1)_D$-gravitational anomaly.} 
Since the defects are expected to be trivially charged under the $SU(4)_R$
R-symmetry, they will not contribute to the
$c_2(\mathcal{S}_6^+)c_1(\mathcal{L}_D)$ anomaly. This provides further
verification that the contribution to this anomaly from the bulk
$\mathcal{N}=4$ vector multiplet, as in section \ref{sec:from4dspec}, is not
modified by interaction terms on the Coulomb branch.
We will verify in section \ref{sec:onC} that
  this precise defect contribution to the anomaly is replicated via holography
  and spectrum calculations when the $\mathcal{N}=4$ theory is further reduced
to 2d on a complex curve $\Sigma$.

\subsection{Class F with Torus-Fibers: Duality Defects}\label{sec:NonUni}

From our analysis we have determined that, regardless of the particular
elliptic fibration $M_6$ that we choose, or equivalently, what particular
$\tau$-profile we choose on $M_4$, the anomaly polynomial $I_6$ has a
$\mathbb{L}$-universal contribution of the form (\ref{eqn:benandholly}). We would like now
to understand how the terms, that we have written as $\cdots$ in
(\ref{eqn:benandholly}) are dependent on the choice of fibration. 

We may consider smooth elliptic fibrations, $M_6$, where the singular fibers
supported over loci in $M_4$ of real codimension $\leq 2$, are all either $I_0$ or
$I_1$ fibers, in the notation of Kodaira \cite{Kodaira}. In this case the pushforward formula
for the product elliptic fourfold, $Y = M_6 \times Z$ is
\begin{equation}
  \pi_*(1 + c_1(Y) + c_2(Y) + c_3(Y) + c_4(Y)) = 12 c_1(\wL)c(B)(1 +
    \alpha c_1(\wL) + \beta c_1(\wL)^2) \,,
\end{equation}
for some numerical coefficients $\alpha$, $\beta$ which depend on the choice
of fibration. We find using standard resolutions of the elliptic fibration, e.g. as in \cite{Marsano:2011hv, Lawrie:2012gg} using {\tt smooth} \cite{Smooth},
\begin{equation}
  \pi_* \left( -p_2(M_6) + \frac{1}{4}p_1(M_6)^2 \right) = 12 c_1(\wL)
      p_1(M_4) - (24\beta +24\alpha + 12)c_1(\wL)^3 \,.
      \label{koko}
\end{equation}
We can understand the dependence on the particular form of the fibration, as
there will be a different network of defects in the 4d theory depending on the
singular fibers supported above codimension two loci in $M_4$. For example, if
we consider a smooth Weierstrass model $M_6$ then it contains the following
Kodaira singular fibers:
\begin{equation}
  \text{Smooth Weierstrass:} \quad \begin{cases}
    \,\,I_1 \,\text{ fibers over }\, 12 c_1(\wL) \cr
    \,\,II  \,\text{ fibers over }\, 24 c_1(\wL)^2    
  \end{cases} \,,
\end{equation}
and in this case we have $\alpha = -6$ and $\beta = 36$, so that from (\ref{koko}) we read off
\be
k_{DDD} = -  r_G \frac{183}{2}\, .
\ee
In the case where we consider $M_6$ to be smooth, but to have a
rank one Mordell--Weil group, i.e.~two sections\footnote{The group of sections, i.e.~maps
from the base to the fiber, of an elliptic fibration is the Mordell--Weil
group. The generic Weierstrass model has one section, which corresponds to one marked
point and is the origin of the elliptic curve. The case of rank one
Mordell--Weil group corresponds to two marked points.}, we have the following
singular fiber structure
\begin{equation}
  \text{Smooth rank one Mordell--Weil group:} \quad \begin{cases}
    \,\,I_1 \,\text{ fibers over }\, 12 c_1(\wL) \cr
    \,\,II  \,\text{ fibers over }\, 48 c_1(\wL)^2 \cr 
    \,\,I_2  \,\text{ fibers over }\, 24 c_1(\wL)^2    
  \end{cases} \,.
\end{equation}
We expect that the case with multiple sections will correspond to adding
punctures, with additional data associated to them, which will be discussed in
the future.  Specifically, and as expected, the defect spectrum of
these class F theories depends on the particular choice
of $\tau$-profile over the 4d space-time -- we see that the $k_{DDD}$
coefficient is sensitive to this, through the parameters $\alpha$ and $\beta$.

Allowing for singular fibers in complex codimension one in the base of the elliptic fibration, 
there are additional contributions to the anomaly polynomial. Singular fibers 
correspond to defects and their Kodaira fiber type determines the additional flavor symmetry. 
The pushforwards of $I_8$ can be computed by first resolving the singularities and using the intersection 
ring of the resolved elliptic fibration. This is e.g.~easily implemented by realizing the Weierstrass model as a Tate model \cite{Bershadsky:1996nh, Katz:2011qp}. 
We use standard resolution methods with conventions as in \cite{Marsano:2011hv, Lawrie:2012gg, Hayashi:2013lra, Hayashi:2014kca, Krause:2011xj,Esole:2011sm}. The pushforward for all Kodaira fibers of type $I_n$ over $S_{G_F}$ (a component of the discriminant, which is a divisor in the base of the fibration), where the flavor symmetry  is $G_F= SU(n)$ is 
\be\ba
& \pi_* \left( - p_2(M_6) + {1\over 4} p_1(M_6)^2 \right)\cr
& \qquad \qquad =  -732 c_1(\wL)^3 + 12 \,\mathfrak{a}_2 \, c_1(\wL)^2
S_{G_F}+c_1(\wL) \left(12 p_1(M_4) - 6 \, \mathfrak{a}_1\, S_{G_F}^2\right) +  12 \, \mathfrak{a}_0 \, S_{G_F}^3  \,,
\ea\ee
where the coefficients for the low values of $n$ for $SU(n)$, realized in terms of $I_n$ Kodaira fibers, are
\be
\begin{array}{c||c|c|c|c|c|c}
n & 2&3&4&5&6&7 \cr \hline
\mathfrak{a}_0 & 1 &4& 10& 20& 35& 56\cr 
\mathfrak{a}_1 &28 & 64& 112& 175& 251& 342\cr 
\mathfrak{a}_2 &  49 & 76& 100& 125& 149& 174 
\end{array}
\ee
Again, these were computed from the explicit resolution of the Tate model.
Note that $\mathfrak{a}_0= n(n + 1)(n + 2)/6$. For the flavor group $G_F$ corresponding to the exceptional groups we find 
\be
\ba
SO(10)= E_5:\qquad &  \mathfrak{a}_0 = 20  \,,\qquad  \mathfrak{a}_1 = 176  \,,\qquad  \mathfrak{a}_2 =  126   \cr 
E_6: \qquad &  \mathfrak{a}_0 = 21 \,,\qquad  \mathfrak{a}_1 = 183  \,,\qquad  \mathfrak{a}_2 =  129 \cr 
E_7:\qquad &  \mathfrak{a}_0 = 24  \,,\qquad  \mathfrak{a}_1 = 200  \,,\qquad  \mathfrak{a}_2 =   135 \cr 
E_8 :\qquad & \mathfrak{a}_0 = 40  \,,\qquad  \mathfrak{a}_1 =  280 \,,\qquad  \mathfrak{a}_2 =   160 \,.
\ea
\ee
Note that these seem to only depend on the flavor symmetry group, as one can check by comparing type $III$ and $I_2$, which both have $SU(2)$. It would indeed be interesting to prove a general expression for any flavor symmetry for the non-universal terms.\footnote{We thank Dave Morrison for discussions on this point.}

In summary, we observed that the duality defects of class F with torus-fibers are classified in terms of a Kodaira classification of singular fibers, which determine the flavor symmetry that the defects carry. Note that duality defects can intersect at points, where the singularity, and thereby the flavor symmetry, enhance, as was shown in \cite{Assel:2016wcr}. 

\subsection{Anomaly for Class F with Genus $g>1$}
\label{sec:ClassFNU}

The theories of class F with fiber $C_g$ for $g > 1$ can be defined by considering the 6d $(2,0)$ theory
dimensionally reduced along the fiber $C_g$, to a 4d theory on $M_4$. As the fibers now have non-trivial curvature the
reduction must be accompanied by a topological twist, as in the standard class
S construction, and in constrast to the $T^2$ fibration discussed in section
\ref{sec:M5T2}.
We consider plane curve fibrations as introduced in section \ref{sec:Fibg}, which can be constructed by taking
a bundle $\mathcal{J}$ on $M_4$ and defining the  fibration over $M_4$
via the projectivisation
\begin{equation}
  \mathbb{P}\mathcal{E} = \mathbb{P}(\mathcal{O} \oplus \mathcal{J}^a \oplus
  \mathcal{J}^b) \,.
\end{equation}
$M_6$ is then  the hypersurface in $\mathbb{P}\mathcal{E}$ of class $dH + e
c_1(\mathcal{J})$, where $H$ is the hyperplane class of the projective
fibration and is then fibered by genus $g$ curves $C_g$.

To consider the 6d $(2,0)$ SCFT on $M_6$ we must topologically twist to cancel
off the curvature in the fibral curves. We decompose the 6d R-symmetry as
\begin{equation}
  \begin{aligned}
    SO(5) &\rightarrow SU(2)_R \times U(1)_R \,,
  \end{aligned}
\end{equation}
and we identify the Chern roots of the $SU(2)_R$ and $U(1)_R$ factors,
respectively $\alpha$ and $r$, in terms of the $SO(5)$ bundle $N_5$ as
\begin{equation}
  n_1 = 2 r - (c_1(M_6) - c_1(M_4) + (e - a - b)c_1(\mathcal{J})) \,, \quad n_2 = 2 \alpha \,.
\end{equation}
We have incorporated the twist of the holonomy along the fibral curve with the 
$U(1)_R$ via the shift of the $U(1)_R$ Chern root by 
\begin{equation}
  c_1(M_6) - c_1(M_4) + (e - a - b)c_1(\mathcal{J}) \,.
\end{equation}
We are twisting to cancel off the curvature of $M_6$ that is transverse to the
embedded $M_4$, which involves shifting by the Chern root of the relative 
tangent bundle, $c_1(T_{M_6/M_4})$, which can be generally expressed as\footnote{We are neglecting here the contributions from the singular fibers, which are immaterial for the purpose of defining the twist.}
\begin{equation}\label{eqn:Cgtwist}
 c_1(M_6) - c_1(M_4)=   c_1(T_{M_6/M_4}) = -(e - a - b)\pi^* c_1(\mathcal{J}) + {c_1(\mathcal{R})} \,,
\end{equation}
where $\mathcal{R} $ {is defined through} the tensor product bundle in
(\ref{eqn:reltgt}). This bundle $\mathcal{R}$ is trivial in the case of a
genus one fibration, when there is no curvature that is required to be
cancelled off transverse to $M_4$. When the $C_g$ fibration is trivial the
bundle $\mathcal{J}$ is trivial, and so $\mathcal{R}$ is simply related to the
tangent bundle to the curve, $T_{C_g}$, which must be twisted with to
compactify on the $C_g$.  As such we must twist only by $\mathcal{R}$, and
thus the twist is given by shifting the Chern root of the $U(1)_R$ by
(\ref{eqn:Cgtwist}).  The anomaly polynomial of the 6d theory, after shifting
the Chern roots as above, can then be written as
\begin{equation}
  \begin{aligned}
    I_8 &= -\frac{d_G h_G^\vee}{6}c_2(R) \left( c_1(M_6) - c_1(M_4) +  (e - a - b)c_1(\mathcal{J}) + 2 c_1(R)
    \right)^2 \cr 
    &\quad + \frac{r_G}{48} \left[ -4 c_2(R) \left( c_1(M_4) - c_1(M_6)+  (e - a - b)c_1(\mathcal{J}) + 2
    c_1(R)\right)^2 \right. \cr
    &\quad \left.  + \frac{1}{4} \left( - 4 c_2(R) + \left( c_1(M_4) - c_1(M_6) +  (e - a - b)c_1(\mathcal{J})+ 2
    c_1(R) \right)^2 - p_1(M_6) \right)^2 - p_2(M_6) \right] \,,
  \end{aligned}
\end{equation}
where we have suppressed the pullbacks on the forms $c_1(R)$, $c_2(R)$,
and $c_1(M_4)$.

When $M_6$ is the hypersurface  fibration defined above we can
pushforward this eight-form anomaly polynomial onto the base, $M_4$, using the
methods described in \cite{MR3094035}. One then finds
\begin{equation}\label{eqn:classfgap}
  \begin{aligned}
    I_6 = \pi_* I_8 &= \kappa_1 c_2(R)c_1(R) + \kappa_2 c_1(R)^3 + \kappa_3
    c_1(R)p_1(M_4) \cr
    &\quad + \kappa_4 c_1(\mathcal{J})c_2(R) +
    \kappa_5c_1(\mathcal{J})c_1(R)^2 + \kappa_6 c_1(\mathcal{J})p_1(M_4) \cr
    &\quad + \kappa_7 c_1(\mathcal{J})^2c_1(R) + \kappa_8 c_1(\mathcal{J})^3 \,,
  \end{aligned}
\end{equation}
where 
\begin{eqnarray}
    \kappa_1 &= &- \left(\frac{4}{3}d_Gh_G^\vee + r_G\right) (g-1) \cr
    \kappa_2 &= &\frac{1}{3} r_G (g-1) \cr
    \kappa_3 &= &- \frac{1}{12} r_G (g-1)  \cr
    \kappa_4 &=& -\frac{1}{6} (d-3)^2 d_Gh_G^\vee (d (a+b)-e)+ \frac{1}{12} r_G (3 ((d-3) d+4) e-d (d (2 d-9)+13) (a+b)) \cr
    \kappa_5 &=& \frac{1}{12} r_G (3 ((d-3) d+4) e-d (d (2 d-9)+13) (a+b)) \cr
    \kappa_6 &= &\frac{1}{48} r_G ((d (d+3)-6) e-d (3 d-5) (a+b)) \cr
    \kappa_7 &=& \frac{1}{24} (d-3) r_G \left[d \left(a^2 (2 (d-3) d+7)+2 a b ((d-3) d+5)+b^2 (2 (d-3) d+7)\right)\right. \cr
  && \quad \left.  -2 (d (2 d-3)+4) e (a+b)+3 d e^2\right]\cr 
    \kappa_8 &=& \frac{1}{48} r_G \left[
      d^2(a+b)\left(
      6 d^2 \left(a^2+b^2\right)-18 d \left(a^2+b^2\right)+3 \left(7 a^2+2 a b+7 b^2\right)\right)
      \right. \cr
    & &\left.
    +e   \left(-2 d^4 \left(a^2+a b+b^2\right)-12 d^3 \left(a^2+a b+b^2\right)+d^2 \left(29 a^2+26 a b+29 b^2\right)
   \right.    \right. \cr
     &&\left. \left.
    -3 d \left(7 a^2+10 a
   b+7 b^2\right)+12 (a+b)^2\right)-d (a+b) \left(11 a^2+10 a b+11 b^2\right)
    \right. \cr
    &&\left. +d e^2 (a + b) \left(6 d^2 +9 d -15 
   \right)+\left(-7 d^2-3 d+6\right) e^3 \right]\,.
\end{eqnarray}
One specialization is to the case of $C= T^2$, for the smooth Weierstrass model
where $\omega = (2,3,3,6)$, which then matches the results in sections  \ref{sec:M5T2} and \ref{sec:NonUni}. 

{We can  also {specialize  to} class S theories by studying the theory where
the fibration is trivial, $\mathcal{J} = \mathcal{O}$, and $a=b=e=0$.} 
In this case {the only non-zero coefficients are} $\kappa_i$ for $i = 1, \cdots 3$, and the anomaly
polynomial for the 4d theory is 
\begin{equation}
  I_6^{\text{class S}} = (g-1) \left[- \left(\frac{4}{3}d_Gh_G^\vee + r_G\right) 
  c_2(R)c_1(R) + \frac{1}{3} r_G c_1(R)^3 - \frac{1}{12} r_G c_1(R)p_1(M_4)
\right] \,.
\end{equation}
This matches the class S anomaly polynomials as calculated in \cite{Alday:2009qq}. 
The non-trivial features of the fibration enter through the coefficients
$(a,b,d,e)$, moreso than just through the genus $g(d)$, which is the only
relevant information when the fibration is trivial.


\section{Class F on $\mathbb{R}^{1,1}\times \Sigma^\tau$}
\label{sec:onC}

As an interesting application and cross-check of the results on class F theories  for $C= T^2$ and their anomalies, we consider 
dimensional reductions to 2d SCFTs. This is interesting for two reasons: first of all, the resulting 2d theories have been studied recently \cite{Haghighat:2015ega, Lawrie:2016axq}, in relation to strings in 6d, and the anomaly polynomial of the resulting  2d  theory was determined in \cite{Lawrie:2016axq} from field theory and for certain cases holographically in \cite{Couzens:2017way}.

The setup will be a class F $T[T^2, \mathcal{F}]$, twisted dimensionally reduced on a curve $\Sigma^\tau$ over which the coupling varies, i.e.~from a 6d point of view, 
\be\label{M6Sigma}
M_6= \mathbb{R}^{1,1} \times \left(T^2\hookrightarrow \mathcal{S}_4 \rightarrow \Sigma^\tau \right) \,,
\ee
and the class F theories we consider are obtained by reducing along $T^2$, and $\mathcal{S}_4$ is the elliptic surface with base $\Sigma^\tau$. The
2d SCFTs are then obtained by further dimensional reduction along $\Sigma$.  A
brane-realization of this is given in terms of D3-branes in an F-theory
background, given in terms of a Weierstrass  elliptic fibration $\pi:Y
\rightarrow B$, where $Y$ is an elliptic Calabi--Yau and $B$ the K\"ahler base
manifold.  In this case $\Sigma^\tau \subset B$, and the varying coupling is
induced from the axio-dilaton variation in F-theory/Type IIB {(for recent reviews
on F-theory see \cite{Heckman:2018jxk, Weigand:2018rez}).}

The topological twist that is required for this reduction is referred to as the
topological duality twist
\cite{Martucci:2014ema,Haghighat:2015ega,Assel:2016wcr,Lawrie:2016axq} and
combines the local Lorentz symmetry on $\Sigma$, $U(1)_\Sigma$, and the
$U(1)_D$ with a $U(1)_R$ inside the $SU(4)_R$ R-symmetry
\begin{equation}\label{eqn:2twists}
  T^{\text{twist}}_\Sigma = T_\Sigma \pm T_R  \,, \quad T^{\text{twist}}_D = T_D \pm T_R \,.
\end{equation}
The amount of supersymmetry that is retained in 2d depends on the embedding of $\Sigma^\tau$ into the ambient elliptic CY $n$-fold. There are essentially three interesting cases to consider: $\Sigma^\tau\subset \CY_3$, which results in strings in 6d, which have $(0,4)$ supersymmetry, $\Sigma^\tau \subset \CY_4$, which results in $(0,2)$ and finally embedding into an elliptic K3, which 
gives rise to $(0,8)$ supersymmetric strings. We now determine the anomaly polynomials starting with class F, and compare them with the known $I_4$ of these supersymmetric strings.

\subsection{$\Sigma$ in $\CY_{3}$: Strings in 6d}\label{sec:CY3bdl}

In this subsection we will consider class F for  $C= T^2$, i.e.
$\mathcal{N}=4$ SYM with varying coupling, with space-time given by
(\ref{M6Sigma}), where $\Sigma \subset B_2 \subset \CY_3$, is a curve in the
base of an elliptic Calabi-Yau three-fold $\pi : Y \rightarrow B$. {For
simplicity we consider gauge group $G=U(N)$.} First we consider the reduction
of the R-symmetry group that is induced by placing the 4d $\mathcal{N} = 4$
SYM theory on a curve inside of a  $B_2
\subset \CY_3$,
\begin{equation}
  SO(6)_R \rightarrow SU(2)_+ \times SU(2)_- \times U(1) \,.
\end{equation}
For the anomaly we are interested in the decomposition of the spin
representations of $SO(6)$ under this reduction
\begin{equation}
  {\bf 4} \rightarrow ({\bf 2,1})_1 \oplus ({\bf 1,2})_{-1} \,.
\end{equation}
In terms of the bundle $\mathcal{S}_6^+$ associated to the ${\bf 4}$
representation this decomposition corresponds to the bundle decomposition
\begin{equation}
  \mathcal{S}_6^+ \rightarrow N^+ \otimes U \oplus N^- \otimes \overline{U}
  \,,
\end{equation}
where $N^\pm$ are the complex vector bundles associated to the fundmental
representations of the $SU(2)_\pm$, and $U$ is the bundle associated to the
charge $+1$ representation of the $U(1)$. Subsequently, the Chern classes become
\begin{equation}
  \begin{aligned}
    c_2(\mathcal{S}_6^+) &= c_2(N^+) + c_2(N^-) - 2 c_1(U)^2 \cr
    c_3(\mathcal{S}_6^+) &= 2 c_1(U) \left( c_2(N^-) - c_2(N^+) \right) \,.
  \end{aligned}
\end{equation}

To determine the anomaly polynomial of the 2d theory on $\mathbb{R}^{1,1}$, we integrate 
 (\ref{eqn:POLY}) over the curve $\Sigma^\tau$ where the
$U(1)$ from the decomposition of the R-symmetry is topologically twisted with
the holonomy of the curve and the $U(1)_D$ as in (\ref{eqn:2twists}). This results in 
\begin{equation}\label{eqn:I4d}
  \begin{aligned}
  I_4 &= \int_\Sigma I_6 \cr 
  &= N^2 \left( c_2(N^-) - c_2(N^+) \right) \int_\Sigma c_1(U) 
   -
   \frac{1}{2} N \left( c_2(N^+) + c_2(N^-) - \frac{1}{2}
   p_1(TM_{2})\right) \int_\Sigma c_1(\mathcal{L}_D) \,.
\end{aligned}
\end{equation}
The topological duality twist involves fixing the first Chern class of $U$ as an appropriate
linear combination of the Chern roots, $t$ and $\delta$, of $T_\Sigma$ and
$\mathcal{L}_D$ respectively. This is 
\begin{equation}\label{eqn:Utw}
  c_1(U) = - \frac{1}{2} t + \frac{1}{2} \delta \,, 
\end{equation}
and thus, by adjunction,
\begin{equation}
  \int_\Sigma c_1(U) = \frac{1}{2} \Sigma \cdot \Sigma \,,
\end{equation}
where we used that the duality bundle
\begin{equation}
  \mathcal{L}_D = \mathcal{O}(-K_B) \,,
\end{equation}
is the canonical bundle of the base of the elliptic Calabi--Yau.
In this way we conclude that the 2d anomaly polynomial is
\begin{equation}\label{eqn:I4conc}
  \begin{aligned}
    I_4 &= c_2(N^-) \left( \frac{1}{2}N^2 \Sigma \cdot \Sigma - \frac{1}{2} N c_1(B)
    \cdot \Sigma \right) \cr &\quad + c_2(N^+)  \left( - \frac{1}{2}N^2 \Sigma \cdot \Sigma -
    \frac{1}{2} N c_1(B)\cdot \Sigma \right) \cr &\quad + \frac{1}{4} N c_1(B)
    \cdot \Sigma p_1(TM_2) \,.
  \end{aligned}
\end{equation}
This matches with the one determined from the spectrum of
the compactification \cite{Haghighat:2015ega,Lawrie:2016axq} as summarised in
appendix \ref{app:CY3spec}.

The general anomaly polynomial for a 2d $(0,4)$ theory with a collection of
$SU(2)_{F}$ flavor symmetries is
\begin{equation}
  I_4 = - \sum_F k_{F} c_2(F) - \frac{1}{24} k p_1(TM_2) \,.
\end{equation}
As such we can read the 't Hooft coefficients off from (\ref{eqn:I4conc})
\begin{equation}
  \begin{aligned}
    k_R& = k_+ = \frac{1}{2} N^2 \Sigma \cdot \Sigma + \frac{1}{2} N c_1(B) \cdot \Sigma \cr
    k_L &= k_- = -\frac{1}{2} N^2 \Sigma \cdot \Sigma + \frac{1}{2} N c_1(B) \cdot \Sigma \cr
    k &= c_R - c_L = -6 N c_1(B) \cdot \Sigma \,.
  \end{aligned}
\end{equation}
Since $SU(2)_R = SU(2)_+$ is the superconformal R-symmetry, the superconformal
algebra allows us to determine the right-moving central charge
\begin{equation}
  c_R = 6 k_R = 3 N^2 \Sigma \cdot \Sigma + 3 N c_1(B) \cdot \Sigma \,,
\end{equation}
and this matches
 with the central charge as given
in \cite{Lawrie:2016axq} and checked holographically in \cite{Couzens:2017way}.

Further we can consider the subcase when the coupling $\tau$ is constant
over the curve, $\Sigma$. In this case $B$ is a K3 manifold, and thus
we consider the compactification of
constant coupling $\mathcal{N} = 4$ SYM on a curve inside of a K3. As is
well-known this gives rise to a sigma model into the Hitchin moduli space
\cite{Bershadsky:1995vm}. In this case the duality bundle is trivial, and thus
\begin{equation}
  \delta = c_1(\mathcal{L}_D) = 0 \,,
\end{equation}
and one can read off from (\ref{eqn:I4d}) that
\begin{equation}
  I_4 = N^2 \left( c_2(N^-) - c_2(N^+)\right) \int_\Sigma c_1(U) \,.
\end{equation}
One does this twist as in (\ref{eqn:Utw}) 
\begin{equation}
  \int_\Sigma c_1(U) = \int_\Sigma - \frac{1}{2} t = g - 1 \,,
\end{equation}
and thus
\begin{equation}
  I_4 =  N^2 ( g - 1 )  \left( c_2(N^-) - c_2(N^+)\right) \,.
\end{equation}

Since $SU(2)_+$ is the right-moving R-symmetry of the resulting $\mathcal{N} =
(4,4)$ theory we can compute the central charge directly: 
\begin{equation}
  c_R = 6 k_R =  6 N^2 (g - 1) \,.
\end{equation}

\subsection{$\Sigma$ in $\CY_{4}$: Strings in 4d}\label{sec:VCY4}

Embedding $\Sigma^\tau$ into an elliptically fibered Calabi--Yau fourfold, as a curve in the base of the fibration gives rise to $(0,2)$ SCFTs. To obtain the anomaly polynomial here, we 
 we integrate  (\ref{eqn:POLY}) resulting in 
\begin{equation}\label{eqn:CY4I4}
  \begin{aligned}
    I_4 = \int_\Sigma I_6 &= - \frac{1}{2} N (- 2 c_1(U)^2 )\int_\Sigma
    c_1(\mathcal{L}_D) + \frac{1}{4} N p_1(TM_2) \int_\Sigma c_1(\mathcal{L}_D)
    \cr
    &= N c_1(B)\cdot \Sigma c_1(U)^2 + \frac{1}{4} N c_1(B)
    \cdot \Sigma p_1(\mathcal{L}_D) \,.
  \end{aligned}
\end{equation}
This anomaly polynomial matches expectations from the direct computation of
the anomaly from the spectrum as reiterated in appendix \ref{app:CY4spec}.
Using the general form of the anomaly polynomial for a $(0,2)$ theory, with a
single $U(1)_F$ flavor symmetry
\begin{equation}
  I_4 = \frac{1}{2} k_{FF} c_1(F)^2 - \frac{1}{24} k p_1(TM_2) \,,
\end{equation}
we can read off the 't Hooft anomalies coefficients from (\ref{eqn:CY4I4})
\begin{equation}\label{eqn:CY4tHooft}
  \begin{aligned}
    k_{UU} &= 2 N c_1(B) \cdot \Sigma \cr
    k &= c_R - c_L = -6 N c_1(B) \cdot \Sigma \,.
  \end{aligned}
\end{equation}
We note that it is not possible in this case to determine the central charges,
as the superconformal R-symmetry can be a linear combination of the $U(1)$ 
global symmetries of the theory, including those that may arise in the defect
sector, as determined by c-extremisation \cite{Benini:2012cz}.

\subsection{$\Sigma$ in K3: Strings in 8d}

Finally, we consider the case when class F for $C= T^2$ is reduced on the base $\mathbb{P}^1$ of an elliptic K3, which gives rise to an $\mathcal{N}=(0,8)$ supersymmetric theory in 2d.
These are strings from wrapped D3-branes in an F-theory compactification on K3 to 8d. The anomaly polynomial for $N$ such strings has not been determined thus far and we briefly derive this here.  
The setup can be studied from 6d on a
surface $S=K3$ and integrating the anomaly polynomial $I_8$ of the 6d
$(2,0)$ theory over $S$. Again, we specialise to the $U(N)$ case where 
$d_G = N^2$ and $r_G = N$.
As the curve on which the theory of class F is reduced is exactly the
$\mathbb{P}^1$ base of the elliptic
fibration, the normal bundle remains ${N}_6$, characterizing the
normal directions of the string in 8d; as such the ${N}_5$ bundle appearing in
the 6d anomaly polynomial lifts directly to a subbundle of the $SO(6)$ bundle
${N}_6$. The only contribution of the anomaly
polynomial is 
\be
I_4 = \int_S I_8 = -{N\over 48} \left( -\frac{1}{2}(p_1 ({N}_5) +
p_1(M_2)) \right) \int_S p_1(TS) 
\ee
For a K3 surface $p_1 (TS)= - 2 c_2 (TS) = - 48$ so that 
\be\label{eqn:K3res}
I_4 = {N\over 2} p_1({N}_5) + \frac{N}{2} p_1(M_2) \,.
\ee

We now wish to compare this anomaly polynomial to the anomaly polynomial
determined by integrating (\ref{eqn:POLY}) over the $\mathbb{P}^1$ base of the
elliptic K3. As there is no twist with the R-symmetry, just a twist between
the holonomy of the curve and the $U(1)_D$ duality symmetry, only the
$c_1(\mathcal{L}_D)$ terms are relevant for the integrated anomaly polynomial.
We find
\begin{equation}
  \begin{aligned}
    I_4 = \int_{\mathbb{P}^1} I_6 &= - \frac{1}{2} N c_2(\mathcal{S}_6^+)
    \int_{\mathbb{P}^1} c_1(\mathcal{L}_D) - \frac{1}{24} (-6 N) p_1(M_2) \int_\Sigma
    c_1(\mathcal{L}_D) \cr
    &= - N c_2(\mathcal{S}_6^+) + \frac{1}{2} N p_1(M_2) \,,
  \end{aligned}
\end{equation}
where we have used that, for a K3 fibration,
\begin{equation}
  \int_{\mathbb{P}^1} c_1(\mathcal{L}_D) = 2 \,.
\end{equation}
It is clear that this matches (\ref{eqn:K3res}) when we use the relation
between the characteristic classes of the $SO(6)$ bundle and its $SO(5)$
subbundle, (\ref{eqn:rosieandjim}).
We can see that for $N=1$ this matches the anomaly polynomial for the
spectrum, as deduced in \cite{Lawrie:2016axq} and summarised in appendix
\ref{app:K3spec}.

\subsection{Class F with $g > 1$ Fibers on $\Sigma$}

In this subsection we consider the anomaly polynomials for the genus $g$
class F theories obtained in (\ref{eqn:classfgap}), and further integrate them along $\Sigma^\tau$,
where
\begin{equation}
  M_4 = M_2 \times \Sigma^\tau \,.
\end{equation}
Again only along $\Sigma$ do we have space-time dependent variation of the  coupling of the 4d class F theory. 
We do the topological duality twisting using the $U(1)_R$ R-symmetry as
\begin{equation}
  c_1(R) = c_1(R^\prime) - \frac{1}{2} ( c_1(\Sigma) + c_1(\mathcal{J}) ) \,.
\end{equation}
Thus we find
\begin{equation}
  \begin{aligned}
    \int_\Sigma I_6 &= p_1(M_2) \left[ \kappa_6 \,\text{deg}(\mathcal{J}) +
      \kappa_3\left( (g_\Sigma - 1) - \frac{1}{2}
      \,\text{deg}(\mathcal{J})\right)
    \right] \cr
    &\quad + c_2(R) \left[ \kappa_4 \,\text{deg}(\mathcal{J}) + \kappa_1\left(
    (g_\Sigma - 1) - \frac{1}{2} \,\text{deg}(\mathcal{J})\right) \right] \cr
    &\quad + c_1(R^\prime)^2 \left[\kappa_5 \,\text{deg}(\mathcal{J}) +
      3\kappa_2\left( (g_\Sigma - 1) - \frac{1}{2}
      \,\text{deg}(\mathcal{J})\right) 
    \right] \,.
  \end{aligned}
\end{equation}
Note that the $SU(2)_R$ is not the superconformal R-symmetry in the IR \cite{Witten:1997yu}.
Furthermore, if we take the class S limit, $\text{deg}(\mathcal{J}) = 0$, then
we find
\begin{equation}
  \begin{aligned}
    \int_\Sigma I_6 &= (g_C - 1)(g_\Sigma - 1)\left(-\frac{1}{12}r_G p_1(M_2)
    + r_G c_1(R^\prime)^2 - \left(\frac{4}{3}d_Gh_G^\vee + r_G\right)c_2(R)
  \right) \,.
  \end{aligned}
\end{equation}
{A  check that may be interesting} to perform is to compare this with the direct reduction of the 6d $(2,0)$ theory on a genus $g$-fibered surface. 

\section{Extensions}
\label{sec:Extensions}

So far we focused on theories of class F, which arise from 6d $(2,0)$ on curve-fibrations. 
Clearly many extensions of this idea are possible. The general, broad principle is as follows: 
consider a supersymmetric theory in $D$ dimensions. Supersymmetry retaining compactifications on complex curves $C$ can be achieved by topologically twisting the theory, i.e.~twisting $U(1)_C$  with a $U(1)_R$ subgroup of the R-symmetry group. For $C= T^2$ the twist is trivial (although one could twist with other global non-R symmetries in this case). 
The duality group of the $D-2$ dimensional theory is usually related to the mapping class group of $C$. 
In any such situation we propose that replacing the direct product with a  $C$-fibration $\pi$ yields a $D-2$ dimensional theory, with duality defects (and thus space-time varying coupling). The anomaly polynomial  can be obtained by pushforward of the $I_{D+2}$ anomaly polynomial
\be
\pi_* I_{D+2} = I_D \,.
\ee
The fibration-related corrections to $I_D$ will depend on a line-bundle that is identified, analogous to the $U(1)_D$ bundle $\mathcal{L}_D$ or geometrically a generalization of the Weierstrass bundle $\mathbb{L}$. 

In this section we illustrate this  with four extensions: first, the non-supersymmetric case, which is in fact far simpler, starting with the 6d self-dual tensor and the class F analog of 4d electromagnetism.
The second extension is to the class of 6d $(1,0)$ SCFTs corresponding to conformal matter theories, which when reduced on a $T^2$ result in 4d $\mathcal{N}=2$ class S theories of specific type. 
Our methods for the pushforward of the anomaly polynomial apply also to this
case and provide a {variant of class F}. Thirdly, class S on an elliptic
fibration which results in 2d theories with space-time varying coupling, and
finally {a large class of} 4d $\mathcal{N}=1$ theories {compactified to 2d
with a duality twist}.

\subsection{Non-supersymmetric Setup: Self-dual Tensor and 4d Maxwell Theory}

The simplest extension is to consider the 6d self-dual tensor (without supersymmetry)\footnote{We thank Yuji Tachikawa for suggesting to consider this case.}, which upon reduction on a $T^2$ gives rise to a 4d $U(1)$ gauge theory, i.e.~free Maxwell theory, with coupling $g$ and theta-angle $\theta$ 
\be
S_{\text{Maxwell}} =  \int_{M_4}{1\over g^2}  F\wedge \star F -  \int_{M_4} {i \theta \over 8 \pi^2} F\wedge F \,.
\ee
As 4d $\mathcal{N}=4$ SYM, this has a duality group acting on the complexified coupling that is identified with the complex structure $\tau$ of the $T^2$. 

Generalizing this to varying coupling is relatively straightforward, and
amounts to a truncation to the gauge sector of the analysis in
\cite{Assel:2016wcr} that derives the maximally supersymmetric case of class F
with $U(1)$ gauge group and varying $T^2$, including the structure of 3d walls
and 2d duality defects. We refrain from repeating the analysis here, but
summarize that the 3d walls correspond to mixed Chern-Simons  couplings for
the S-duality walls, and Chern-Simons couplings at level $k$ for $T^k$ walls.
The 3d walls (whose locations indicate the branch-cuts of $\tau$) end on the
duality defects, which are real codimension 2. The chiral theories on the
duality defects are WZW models, with gauge group specified by the Kodaira
singularity type. 

What we determine here is the corrections to the anomaly polynomial. For the
constant $\tau$ case this was discussed recently in \cite{Seiberg:2018ntt}
including generalizations to moduli-space dependent terms. Here we determine
the additional terms due to varying coupling.  Recall for this the computation
of the pushforwards for the similar situation of the 6d $\mathcal{N} = (2,0)$
anomaly polynomial (\ref{eqn:I8ap}).  More generally, one can consider the
universal part of the pushforward 
\begin{equation}
  \pi_* \left( - a p_1(M_6)^2 - b p_2(M_6) \right) = 24 ( 2 a + b )
  c_1(\mathbb{L}) p_1(M_4) + \cdots \,.
\end{equation}
In particular, the anomaly polynomial due to a single self-dual tensor field
has \cite{AlvarezGaume:1983ig} 
\begin{equation}
  a^{\text{ s.d.~tensor}} = - 1/360 \,, \quad b^{\text{ s.d.~tensor}} = 7/360
  \,,
\end{equation}
and thus 
\begin{equation}
  \pi_* I_8^{\text{ s.d.~tensor}} = \frac{1}{3} c_1(\wL) p_1(M_4) + \cdots \,.
\end{equation}
Similarly, one can consider the theory of a single chiral fermion in 6d, for
which the coefficients in the gravitational anomaly are
\begin{equation}
  a^{\text{ fermion}} = -7/5760 \,, \quad b^{\text{ fermion}} = 4/5760 \,,
\end{equation}
and again one can see that
\begin{equation}
  \pi_* I_8^{\text{ fermion}} = - \frac{1}{24} c_1(\wL) p_1(M_4) + \cdots \,.
\end{equation}

The non-universal parts corresponding to the defects are again dependent on
the specific singularity of the fibration. As mentioned earlier these will
correspond to chiral WZW models, localized on 2d defects.

\subsection{6d Conformal Matter Theories and Class F}

One can consider 6d $\mathcal{N} = (1,0)$ SCFTs and {fiber-reduce these along an}
elliptic fibration to 4d.
To illustrate this, we consider the 6d theory of a single M5-brane on
$\mathbb{C}^2/\Gamma_G$ which is known as the $(G, G)$ minimal conformal matter theory
\cite{DelZotto:2014hpa} -- though this method will be applicable to all 6d theories, where the anomaly polynomials have been determined. This is particularly interesting in the current setup, as  it was shown in \cite{Ohmori:2015pua} that the 4d
$\mathcal{N}$ = 2 theory arising as the $T^2$ compactification of the $(G,G)$
minimal conformal matter theory is the same as the class S theory that arises
from compactifying the 6d $(2,0)$ theory of type $G$ on an $S^2$ with two full
punctures and an additional (non-standard) puncture (which can be of simple type in certain examples). Class S theories of this form are known as the generalised bifundamental theories. The fibered reduction should thus have an alternative description in terms of class S theories with space-time varying coupling -- i.e.~class F. 

The anomaly polynomial for the 6d $(1,0)$ theory from $N$ M5-branes probing
the ALE singularity $\mathbb{C}^2/\Gamma_G$ was determined in
\cite{Ohmori:2014kda}. On the generic point of the tensor branch the theory is
a quiver gauge theory with flavor and gauge groups 
\begin{equation}
  [G_0] \times G_1 \times \cdots \times G_{N-1} \times [G_N] \,.
\end{equation}
From such a tensor branch descriptioon it can be determined that 
the anomaly polynomial of the 6d $(G, G)$ conformal matter
theory is\footnote{When $G = A_r$ there is a further $U(1)$ global symmetry,
which we disregard in the following discussion.} 
\begin{equation}
  \begin{aligned}
    I_8^{G} &= \frac{|\Gamma|^2N^3}{24} c_2(R)^2 -
    \frac{N}{48}\left(|\Gamma|(r_G + 1) - 1\right)\left(4c_2(R) +
    p_1(T)\right)c_2(R) \cr
    &\quad + \frac{N}{8}\left(\frac{1}{6}c_2(R)p_1(T) - \frac{1}{6}p_2(T) +
    \frac{1}{24}p_1(T)^2\right) \cr
    &\quad - \frac{N}{8}|\Gamma|c_2(R)(\text{Tr}F_0^2 + \text{Tr}F_N^2) -
    \frac{1}{2}I_8^{\text{vec}}(F_0) -\frac{1}{2}I_8^{\text{vec}}(F_N) \cr 
    &\quad - I_8^{\text{tensor}} - \frac{1}{2N}\left(\frac{1}{4}\text{Tr}F_0^2
    - \frac{1}{4} \text{Tr} F_N^2\right)^2 \,,  
  \end{aligned}
\end{equation}
where $|\Gamma|$ is the order of $\Gamma_G \subset SU(2)$, $F_i$ is the curvature of the
gauge/flavor factor $G_i$, and
\begin{equation}
  \begin{aligned}
    I_8^\text{vec}(F) &= -\frac{1}{48} \bigg( \text{tr}_\text{adj}F^2 +
    d_G c_2(R)\bigg)p_1(T) - \frac{d_G}{5760}\bigg( 7p_1(T)^2 - 4p_2(T)\bigg)
    \cr
    &\quad -\frac{1}{24}\bigg(  \text{tr}_\text{adj}F^4
    + 6 c_2(R) \text{tr}_\text{adj}F^2 + d_Gc_2(R)^2\bigg) \cr
    I_8^\text{tensor} &= \frac{1}{24} c_2(R)^2 + \frac{1}{48}c_2(R) p_1(T)  +
    \frac{1}{5760}\left(23 p_1(T)^2 - 116 p_2(T)\right) \,.
  \end{aligned}
\end{equation}

If we apply the same strategy as in section \ref{sec:M5} to integrate the eight-form
anomaly polynomial along the $T^2$ fiber of a non-trivial elliptic
fibration then we can determine part of the anomaly polynomial of the 4d theory
defined on $B$, where $\tau$ is now allowed to vary over that space-time $B$.
In particular we find that
\begin{equation}\label{eqn:GBpush}
  \begin{aligned}
    I_6^G = \pi_* I_8^G &= \left[ - \frac{N}{48}\left(|\Gamma|(r_G + 1) - 2\right) +
    \frac{(d_G-1)}{48} \right]c_2(R)\pi_*p_1(Y) \cr
    &\quad + \left[\frac{1}{5760} \left(30 N + 7 d_G - 23 \right) \right]
    \pi_*\left(p_1(Y)^2\right) \cr
    &\quad + \left[\frac{1}{5760} \left(-120 N - 4 d_G + 116\right) \right] \pi_*p_2(Y)\cr
    &\quad + \left[\frac{1}{96}\left(h_{G_0}^\vee \text{Tr}F_0^2 +
      h_{G_N}^\vee \text{Tr} F_N^2\right)
    \right] \pi_*p_1(Y) \cr
    &= \left[ \frac{N}{2}\left(|\Gamma|(r_G + 1) - 2\right) -
    \frac{(d_G-1)}{2} \right]c_2(R)c_1(\wL) 
    \cr &\quad 
    + \left[ \frac{N}{4} - \frac{d_G}{24} - \frac{7}{24} \right] c_1(\wL)p_1(B)
    - \left[\frac{1}{4}\left( h_{G_0}^\vee\text{Tr}F_0^2 +
      h_{G_N}^\vee\text{Tr}F_N^2\right)
    \right] c_1(\wL) + \cdots \,. 
  \end{aligned}
\end{equation}
These anomaly coefficients are $\mathbb{L}$-universal, independent of the $\tau$-profile
with which we endow the 4d space-time, and $\cdots$ stands for non-universal
terms. {Universal refering here to} terms that follow from the expressions
for the pushforwards of the combinations of Chern classes in
(\ref{eqn:cnmost2}) and (\ref{eqn:c2c2}). {We emphasize that, as one can
  see from appendix \ref{app:CCs}, the non-unverisal terms involve either
  higher than linear order terms in $c_1(\mathbb{L})$ or else Chern classes of
other bundles related to the singular fibers of the elliptic fibration, and as
such the $\mathbb{L}$-universal terms written above are not corrected by the non-universal
terms.}

This proposal has a nice cross-check by considering further reduction from 4d along a curve $\Sigma$ to a 2d theory with $(0,2)$ supersymmetry. This is equivalently the 6d conformal matter theory on an elliptic surface, which is $C$-fibered over $\Sigma$. 
I.e.~we will now compare the two anomaly polynomials for the 2d theories arising as
the compactification of the 4d $\mathcal{N} = 2$ theory with anomaly
polynomial (\ref{eqn:GBpush}) on a curve $\Sigma$ over which the bundle $\wL$
is supported or alternatively as coming from the $\mathcal{N} = (1,0)$ theory
with anomaly polynomial $I_8^G$ compactified on an elliptic surface
\begin{equation}
  T^2 \hookrightarrow P \rightarrow \Sigma \,.
\end{equation}
In the first instance this involves a notion of topological duality twist,
since $\wL$ is a bundle on the compactification space $\Sigma$, however in the
second construction this is simply a geometric twist cancelling off the
curvature of the elliptic surface.

Let us first consider $I_8^G$ compactified on the elliptic surface $P$. We
decompose the tangent bundle as
\begin{equation}
  TM_6 = TM_2 \oplus TP \,,
\end{equation}
and the $SU(2)$ R-symmetry is decomposed into a $U(1)_R$ symmetry, which will
become the R-symmetry on the resulting 2d $(0,2)$ theory, and which is twisted
with the curvature on $P$:
\begin{equation}
  c_2(R) = -\left(c_1(R) - \frac{1}{2}c_1(TP)\right)^2 \,.
\end{equation}
When integrating we find
\begin{equation}\label{eqn:I8GonP}
  \begin{aligned}
    \int_P I_8^G &= \frac{1}{48} \left[ (N(|\Gamma|(r_G + 1) - 2) - d_G) c_1(R)^2
    + \left(-\frac{N}{2} + \frac{d_G}{12}\right)p_1(M_2) \right. \cr
    &\quad \qquad \left. + \frac{1}{2} \tr_{\text{adj}}F_0^2 + \frac{1}{2} \tr_{\text{adj}}F_N^2
    \right] \int_P p_1(P) \,,
  \end{aligned}
\end{equation}
where 
\begin{equation}
  \int_P p_1(P) = -24 \,\text{deg}(\wL) \,.
\end{equation}
This matches the anomaly polynomial given in \cite{Apruzzi:2016nfr} for
compactifications of 6d SCFTs on, not necessarily elliptic, K\"ahler
four-manifolds.

Alternatively, we can obtain the same theory from reducing the 4d theory  with the anomaly polynomial (\ref{eqn:GBpush}), which has space-time varying coupling, and compactify
on a curve $\Sigma$, above which the $T^2$ varies. We decompose
\begin{equation}
  p_1(M_4) = p_1(M_2) + p_1(\Sigma) \,,
\end{equation}
and the R-symmetry as
\begin{equation}
  c_2(R) = - ( c_1(R) - \varepsilon_1 c_1(\Sigma) - \varepsilon_2 c_1(\wL))^2\,,
\end{equation}
where the $\varepsilon_i$ are the twisting parameters. In fact, these terms
will be irrelevant for the integration of the anomaly polynomial
(\ref{eqn:GBpush}), as there are no mixed terms involving $c_2(R)$ and an
abelian flavor symmetry. Integrating we find
\begin{equation}
  \begin{aligned}
    \int_P I_6^G &= -\frac{1}{2} \left[ (N(|\Gamma|(r_G + 1) - 2) - d_G) c_1(R)^2
    + \left(-\frac{N}{2} + \frac{d_G}{12}\right)p_1(M_2) \right. \cr
    &\qquad \left. + \frac{1}{2} \text{tr}_{\text{adj}}F_0^2 + \frac{1}{2} \tr_{\adj}F_N^2
    \right] \int_\Sigma c_1(\wL) \,,
  \end{aligned}
\end{equation}
where
\begin{equation}
  \int_\Sigma c_1(\wL) = \text{deg}(\wL) \,,
\end{equation}
demonstrates the equivalence with (\ref{eqn:I8GonP}).

The anomaly polynomial of the 
4d theories from any 6d $(1,0)$ SCFT on a $T^2$ (or even general curve) can be discussed 
completely analogously to the above. It would indeed be very interesting to develop other tools, e.g.~
a microscopic description of the duality defects in this case, to study these theories.

\subsection{{4d $\mathcal{N} = 2$ to 2d with Varying Coupling}}

Another generalization, which we would like to illustrate is the case of 4d to 2d reduction, where the 2d theory has a manifest space-time dependent ``coupling''. E.g.~consider an $\mathcal{N} = 2$ SCFT defined on a space-time $M_4$, which is fibered as
\be\label{eqn:T2T2}
T^2 \hookrightarrow M_4 \stackrel{\pi}{\longrightarrow} M_2\,,
\ee
and consider the reduction along the fibral $T^2$. The 4d anomaly polynomial is 
\begin{equation}
  I_6 = 24(a - c) \left( \frac{1}{3}c_1(R)^3 - \frac{1}{12}c_1(R)p_1(M_4)
  \right) - 4(2 a - c) c_1(R) c_2(R) + \cdots \,,
\end{equation}
where the $c_1(R)$ is from the $U(1)_R$ R-symmetry, the $c_2(R)$ from the
$SU(2)_R$ R-symmetry, and the $\cdots$ are terms involving flavor symmetries. 

We now use that this space-time $M_4$ is an elliptic fibration as in
(\ref{eqn:T2T2}), and we integrate over the fiber. To do this we need to know
only 
\begin{equation}
  \pi_* p_1(Y) = -24 c_1(\wL) \,,
\end{equation}
where we, in the usual manner, treat $M_4$ as an effective elliptic threefold
for the purposes of the anomaly. Then we can see that the anomaly polynomial
of the 2d $\mathcal{N} = (4,4)$ theory will contain correction terms
\begin{equation}
  I_4 = \pi_*I_6 + \cdots = 48(a - c)c_1(R)c_1(\wL) + \cdots \,,
\end{equation}
which manifestly depends on the fibration of the $T^2$.

\subsection{4d $\mathcal{N} = 1$ Quiver Theories with Spacetime-Varying Coupling}

\label{n1section}

In this section we will discuss an extension to ${\cal N}=1$ 4d theories that cannot be obtained as reductions of 6d theories. 
In \cite{Couzens:2017nnr} 
it was proposed that these  
 can be promoted to theories with a space-time varying coupling, and part of their anomaly polynomial can be determined, 
 in a sense that we will clarify   below. Specifically,  we consider quiver gauge theories with a gauge group $G=U(N)^\chi$ and chiral multiplets
  transforming in bi-fundamental representations of pairs of $U(N)\times U(N)\subset G$, 
  and a  superpotential.  A defining property of these theories is that their Abelian vacuum 
  moduli space is a Calabi-Yau threefold conical singularity $\mathbb{X}=C(Y)$.  In the large $N$ limit, they
  are conjectured to flow to interacting SCFTs in the IR, with gauge group $SG=SU(N)^\chi$, while the $U(1)$ factors 
  decouple and remain as $\chi-1$ global (possibly anomalous) baryonic symmetries  and one overall ``center of mass''.
 
 The holographic duals of these theories are type IIB supergravity solutions of the form AdS$_5\times Y$ for which
the (constant) axio-dilaton $\tau$ is a modulus. In the field theories this is  part of the \emph{conformal manifold}.
In the class of ``toric''  quiver theories, namely those for which  $\mathbb{X}$ is a toric Calabi-Yau singularity,  one can identify the combination of couplings dual to $\tau$, as we now briefly recall.
In these theories there are $\chi$ complex gauge couplings $\tau_i$, as well as  $L$ complex superpotential couplings $h_k$. The superconformal point corresponds to the vanishing of the corresponding $A=\chi+L$ beta functions\footnote{$A$ is the number of arrows in the quiver, namely the number of bi-fundamental chiral fields.}, however  there are degeneracies, defining a manifold of exactly marginal deformations. Generically, the dimension of this manifold is $b_3(Y)+2$ \cite{Imamura:2007dc}, where $b_3(Y)$ denotes the third Betti number of $Y$, and it coincides with the number of non-anomalous baryonic $U(1)$'s.  The two additional marginal deformations have been identified with a generalization of the beta-deformation (involving $h_k$'s only) and the diagonal combination\footnote{The imaginary part of the gauge couplings $\tau_i'$ are related to $\tau_i$ as 
$\tfrac{1}{g_i^{'2}}=\tfrac{1}{g_i^{2}}-\tfrac{N}{8\pi^2}\log \tfrac{1}{g_i^2}$ \cite{Imamura:2007dc}, and similarly  for the real part.} 
\be
\tau_{\rm{diag}} = \sum_{i \,\in\, \rm{nodes}}  \tau_i' - N \sum_{k \, \in\,  \rm{loops}}  h_k \,.
\ee
It is this marginal coupling that we will promote to vary over space-time, while keeping constant the other $A-1$ couplings, marginal or otherwise. 
While the non-marginal couplings are fixed to some value in the IR theory, and therefore it does not make sense to promote these to vary over space-time, 
in principle, it would be interesting to consider other varying marginal couplings. It is straightforward to see how the single couplings  will vary over space-time: consider the relation 
\begin{equation}
M \boldsymbol{t} = \mathbf{v}~, 
\label{yui}
\end{equation}
where $\boldsymbol{t} ^T=(\tau_1',\dots, \tau_\chi',h_1,\dots,h_L)$ and $ \mathbf{v}^T=(f,\dots)$ with $f$ a complex 
 function of space-time coordinates and the remaining $A-1$ entries parameterising the other couplings, marginal or not. For simplicity we can tune the remaining $b_3(Y)+1$ marginal couplings to zero, namely we take
 \begin{equation}
 \mathbf{v}^T=(f,0,\dots,0,v_{b_3(Y)+3},\dots,v_{A})~,
 \end{equation}
 so that they will not transform under $SL(2,\mathbb{R})$ or $U(1)_D$. $M$ is an appropriate $A\times A$ constant matrix, whose first row corresponds to the relation $\tau_{\rm{diag}} =f$, that is identified with the axio-dilaton $\tau$. Inverting 
 (\ref{yui}) one gets 
\begin{equation}
t_i = M^{-1}_{i1}f +  c_i ~,
 \end{equation}
where the $c_i$ are fixed constants, so that under  an $SL(2,\mathbb{R})$ transformation these transform as 
\begin{equation}
t_i \to  M^{-1}_{i1} \frac{a f + b}{c f + d} + c_i~,
\label{funnychange}
 \end{equation}
where notice that all gauge and superpotential couplings transform non-trivially.

The  dual type IIB supergravity solutions enjoy exactly the same 
$U(1)_D$ symmetry of the AdS$_5\times S^5$ solution, acting on the axion-dilaton and  this motivates conjecturing that the    ${\cal N}=1$  field theories
possess a bonus $U(1)_D$ symmetry  \cite{Intriligator:1998ig}, analogous to   ${\cal N}=4$  SYM. 
However, differently from the latter,  from (\ref{funnychange}) one can see that  for these ${\cal N}=1$  theories the $U(1)_D$ is not a symmetry of the equations, even for the abelian case.

In  \cite{Couzens:2017nnr}  it has been proposed that these theories possess an anomaly polynomial that extends that of the theories with constant 
$\tau$ to one including the terms 
\begin{equation}
\begin{aligned}
\delta I_{6}^\tau =  &\frac{1}{2}k_{D IJ}c_{1}({\cal L}_D) c_{1}(F_I)
c_{1}(F_J) + {\frac{1}{2}}k_{DD I}c_{1}({\cal L}_D)^2 c_{1}(F_{I}) \cr
 +  &{\frac{1}{6}}k_{DDD}c_{1}({\cal L}_D)^3- \frac{1}{24} k_{D} c_{1}({\cal L}_D) p_{1}(TM_4)\, ,
\end{aligned}
\end{equation}
where  the index $I$ runs over all the global $U(1)$ symmetries, including the $U(1)_R$ R-symmetry.

As anticipated in  \cite{Couzens:2017nnr}, the bulk contribution to 
 the anomaly coefficients $k_{DIJ}$, $k_{DDI}$, $k_{DDD}$, $k_D$ may be computed by considering the theory at a generic point on the Higgs branch, that for these
theories is the symmetric product  Sym$^N \mathbb{X}$. 
However, this is not the complete answer, as it misses the contribution of the defects degrees of freedom. Here we will argue that the methods discussed in the paper may be applied to derive expressions for $k_{DDD}$ and $k_D$ that incorporate the total contribution of the bulk and defects, for the present theories\footnote{The coefficients $k_{DIJ},k_{DDI}$ involve global symmetries that are not visible in 6d, and therefore our methods do not apply to these.}.

The toric quiver gauge theories  can be realised applying  a simple two-step procedure, starting from the ${\cal N}=4$ SYM theory: firstly, we quotient by a discrete group $\Gamma \subset SU(3)$, obtaining an orbifold ${\cal N}=1$ theory. From the D3 branes point of view, this amounts in quotienting the space  transverse to the branes to $\mathbb{C}^3/\Gamma$; secondly, by appropriate Higgsing 
\cite{Feng:2000mi} of the field theory, one can reach an arbitrary toric theory, provided its toric diagram is included in the toric diagram of the orbifold theory.

At a generic point on the Higgs branch, 
the low energy theory is $U(1)^{N-1}$ and comprises $N-1$ vector multiplets and $3N$ chiral multiplets, parameterizing the flat directions on Sym$^N \mathbb{X}$, plus a free abelian vector multiplet  parameterizing the decoupled center of mass.
 {Thus assuming  all the fermions have  $U(1)_D$ charges $q_D=\frac{1}{2}$,  we have}
\begin{equation}
\begin{aligned}
   k_D |_{\rm bulk} &=    \frac{1}{2} \times \left( N-1+3N +1  \right) = 2 N~, \cr
    k_{DDD} |_{\rm bulk} &=    \left(\frac{1}{2}\right)^3  \times  \left( N-1+3N +1  \right) =
    \frac{1}{2} N \,,
  \end{aligned}
\end{equation}
which are  exactly the same contributions of the parent  ${\cal N}=4$  theory.

As the theory flows to an interacting SCFT in the IR, global symmetries may mix with the UV R-symmetry, however, it is natural to assume that the $U(1)_D$ symmetry does not participate to this mixing, so that the total value of $k_D$ will be 
unchanged\footnote{The anomaly polynomial thus obtained captures  the combined contribution of the interacting SCFT and the center of mass. While  the bulk contributions can be easily disentangled, this is not clear for the defect part.} from that of the parent ${\cal N}=4$ theory computed in (\ref{eqn:benandholly}), namely
\begin{equation}
\begin{aligned}
 k_D =  k_D |_{\rm bulk}  + k_D |_{\rm defect} & = - 6N \,.
     \end{aligned}
\end{equation}
{In \cite{Couzens:2017nnr} it was shown that in various examples in which the 4d ${\cal N}=1$ theories are reduced to 2d $(0,2)$ theories,} this value matches, assuming that the center of mass gives a  ${\cal O}(N^0)$ contribution, at leading order in $N$, with the dual holographic computation {in AdS$_3$}.

\section{Conclusions and Outlook}
\label{sec:CO}

We argued for the presence of new terms in the anomaly polynomial of 4d supersymmetric theories in the presence of duality defects, i.e.~in situations where the 
coupling is space-time dependent and undergoes S-duality transformations. 
Theories of this type, obtained by reducing the 6d $(2,0)$ theory along the fiber of a genus $g$ curve fibration,
 have been  {referred to as} theories of class F. It is for such theories, that the additional terms, which depend on the connection of a line bundle associated to a local $U(1)_D$ symmetry, are non-trivial.

In the simplest case, for 4d $\mathcal{N} = 4$ SYM, 
with gauge group $G$, the anomaly polynomial {was shown to have additional terms to the standard $c_3(\mathcal{S}_6^+)$ R-symmetry anomaly }
\begin{equation}
  I_6 = \frac{1}{2} d_G c_3(\mathcal{S}_6^+) - \frac{1}{2} r_G c_2(\mathcal{S}_6^+)
  c_1(\mathcal{L}_D) - \frac{1}{24} 
  (-6 r_G ) c_1(\mathcal{L}_D) p_1(T_4) - \frac{61}{4} r_G c_1(\mathcal{L}_D)^3 \,,
\end{equation}
where $d_G$ is the dimension of $G$, $r_G$ is the rank of $G$. More precisely, as we explained, this is the anomaly for 
class F with $T^2$-fibrations\footnote{More precisely, this expression is valid for elliptic fibrations that are Weierstrass models with one section and no singular fibers with non-abelian flavor symmetry enhancements.}. The additional terms depending on the specific type of duality defects and additional sections/marked points, which were determined in section \ref{sec:NonUni}.
A similar expression was obtained for $g>1$, which {modifies} the class S anomaly polynomial by terms depending on the varying $C_g$ over space-time (\ref{eqn:classfgap}). 

The new terms, which are sourced by the varying coupling, are genuine corrections to the anomaly polynomials of 4d $\mathcal{N}=4$ SYM and class S theories. Another way of putting the result is that these theories can gain additional terms in the anomaly polynomical in the presence of duality defects. The additional $\mathcal{L}_D$ dependent terms have a striking similarity to the ones that appeared recently in \cite{Minasian:2016hoh} for the Type IIB/F-theory effective action, generalizing the $F_D$-dependent terms in \cite{Gaberdiel:1998ui}. We have seen that for certain compactifications of class F all terms are relevant, except the $c_1(\mathcal{L})^3$ term, which much like the $F_D^6$ term in the anomaly polynomial $I_{12}$ of Type IIB \cite{Minasian:2016hoh}, remains to be tested from an alternative computation. 
It would be interesting to see whether a derivation e.g.~from anoamly inflow is possible, generalizing \cite{Kim:2012wc} for D3-branes in the presence of general $(p,q)$ 7-branes.

There are various extensions that we have not considered. 
Clearly an in depth analysis of the class F with higher genus curve fibrations
should succeed this paper, in particular realizing the theories for all genera
and including the data on punctures. It would also be interesting to elucidate
the $U(1)$ associated to the bundle $\mathcal{J}$ from a 4d viewpoint, for genus $g>1$ theories. It seems that mathematically, the description in terms of Lefschetz fibrations is most suited and we hope to return to this. 
However, as we have seen, for the computation of the anomaly a concrete realization of the fibration, in order to compute the pushforward of the $I_8$ form along the fibers, is required.  It  would be interesting to develop this framework to include punctures on the fibral curves $C_{g,n}$, which correspond to sections of the fibration, and extend this to anomaly considerations, i.e.~generalizing the recent analysis in \cite{Bah:2018gwc} in class S to class F.

\subsection*{Acknowledgements}
We thank Fernando Alday, Benjamin Assel, Chris Couzens, Michael Green, Monica
Kang, Heeyeon
Kim, Dave Morrison, Pablo Soler, and Yuji Tachikawa for discussions. CL
thanks IPMU Tokyo for hospitality during the completion of this work. 
The work of CL was partially supported by DFG under
Grant TR33 ``The Dark Universe'' and under GK 1940 ``Particle Phyiscs Beyond the Standard
Model''.  DM is supported by the ERC Starting Grant 304806 ``The gauge/gravity duality and
geometry in string theory''.
SSN's work is supported by the ERC Consolidator Grant 682608 ``Higgs bundles: Supersymmetric Gauge
Theories and Geometry (HIGGSBNDL)''.

\appendix

\section{Characteristic Classes and Anomaly Polynomials}\label{app:CCs}

In this appendix we will collect several expressions for the characteristic
classes that appear in the anomaly polynomials, and discuss the integration of
anomaly polynomials by summarising how to pushforward Chern classes on genus
$g$ fibered eight-manifolds, $Y$, to the base, via adjunction and the
Grothendieck--Riemann--Roch theorem.

\subsection{Summary of Chararacteristic Classes}
  
Let $\mathcal{E}\rightarrow M$
be a \emph{complex} vector bundle of rank $n$. The Chern classes are defined for complex
vector bundles, and will be denoted by $c_k (\mathcal{E})$.  We denote the
Chern roots, the eigenvalues of the curvature form {$\frac{iF}{2\pi}$}, as $r_i$.  Then the
total Chern class is defined as
\be
{c (\mathcal{E}) =   \det \left(1+ \frac{iF}{2\pi} \right)  \, \equiv 1  + c_1 (\mathcal{E}) + c_2 (\mathcal{E}) +\cdots \, ,}
\ee
with 
\be
c_1 (\mathcal{E}) = \sum_{i=1}^n r_i \,,\qquad 
c_2 (\mathcal{E}) = \sum_{i<j}  r_i r_j\,,\qquad 
c_3 (\mathcal{E}) = \sum_{i<j<k}  r_i r_j r_k\,.
\ee
{A useful property of the total Chern class is that for a direct sum of two complex vector bundles we have}
\be
{c ( \mathcal{E}_1\oplus \mathcal{E}_2) = c ( \mathcal{E}_1)\,  c(  \mathcal{E}_2) \, .}
\label{directsum}
\ee
The Chern character on the other hand is defined as
\be
\ch (\mathcal{E}) = \Tr e^{i{\tfrac{F}{2\pi}}} = \sum_{i=1}^n e^{r_i} \,
\equiv ch_0 (\mathcal{E})  + \ch_1 (\mathcal{E}) + \ch_2 (\mathcal{E}) +\cdots
\, ,
\ee
specifically
\be
\ba
\ch_0 (\mathcal{E}) & = n \, ,\cr
\ch_1 (\mathcal{E}) &= c_1 (\mathcal{E}) \, , \cr 
\ch_2 (\mathcal{E}) &= {1\over 2} (c_1 (\mathcal{E})^2 - 2 c_2 (\mathcal{E})) \, , \cr 
\ch_3 (\mathcal{E}) &= {1\over 6} ( c_1 (\mathcal{E})^3 - 3 c_1 (\mathcal{E})c_2 (\mathcal{E}) + 3 c_3 (\mathcal{E}))\,.
\ea
\ee
A useful property of the Chern character is that for a tensor product of two complex vector bundles we have
\be
\ch ( \mathcal{E}_1\otimes \mathcal{E}_2) = \ch ( \mathcal{E}_1)\,  \ch(  \mathcal{E}_2) \, .
\ee

{Let $E \rightarrow M$ be a real \emph{vector} bundle of (even) rank $k$.
The Pontryagin classes are  defined in terms of Chern classes of the complexification  of  $E$, $E^\mathbb{C}=E \otimes \mathbb{C}$, as}
\be
{p_j (E) = (-1)^j c_{2j} (E^\mathbb{C})~.}
\ee
{Equivalently the  total Pontryagin class may be defined in terms of 
the block-diagonal eigenvalues  $\rho_i$ of the  curvature $\frac{F}{2\pi}$ of the connection  on the bundle $E$. Namely}
\be
p(E) = \det \left(1+ \frac{F}{2\pi}\right) {\equiv 1 + p_1(E) + p_2(E) +\dots } = \prod_{i=1}^{[k/2]} (1+ \rho_i^2)\,,
\ee
Specifically 
\be\ba
p_1 (E)& =  \sum_{i=1}^{[k/2]} \rho_i^2 \, \cr 
p_2 (E) &= \sum_{i<j} \rho_i^2 \rho_j^2 \,. 
\ea\ee

Note that we can view the tangent bundle $\omega \equiv TM$ to a manifold $M$ as a complex rank $n$ bundle,
or as a real rank $k=2n$ bundle $\omega_\mathbb{R}$. 
{The Pontryagin  classes are defined for its real form $\omega_\mathbb{R}$, and are  given in terms of the Chern classes of the  complexification}
\be
{\omega_\mathbb{R} \otimes \mathbb{C} = \omega \oplus \bar \omega}\, .
\ee
{Using (\ref{directsum})  then yields the relation  \cite{MR0440554}}
\be
{1- p_1(\omega_\mathbb{R})+ p_2(\omega_\mathbb{R})  +\dots  = \left(  1  + c_1 (\omega) + c_2 (\omega) +\cdots   \right)
 \left(1  - c_1 (\omega) + c_2 (\omega) +\cdots  \right) \, .}
\ee
{Specifically, with a slight abuse of notation,  as standard, we have}
\begin{equation}
  \begin{aligned}
    p_1(TM) &= c_1(TM)^2 - 2 c_2(TM)\, ,  \cr
    p_2(TM) &= c_2(TM)^2 - 2c_1(TM)c_3(TM) + 2 c_4(TM) \,,
  \end{aligned}
\end{equation}
{where on the left hand side of these equations $TM$ is viewed as real rank $2n$ vector bundle and on the right hand side as a complex rank $n$ vector bundle.}

{The $\hat A$-roof class of a manifold, $M$, is defined as}
\begin{equation}
  \widehat{A}(M) = 1 - \frac{1}{24} p_1(TM) + \frac{1}{5760}(-4 p_2(TM) + 7
  p_1(TM)^2) + \cdots \,,
\end{equation}
where $TM$ is the {(real)} tangent bundle to $M$. Similarly the Todd {class}
 of $M$ is
defined as
\begin{equation}
  \td(M) = 1 + \frac{1}{2} c_1(TM) + \frac{1}{12} ( c_2(TM) + c_1(TM)^2) +
  \cdots \, ,
\end{equation}
where $TM$ is the {(complex)} tangent bundle to $M$.

\subsection{Integration of Anomaly Polynomials}

We will consider an eight-manifold, $Y$, equipped with a fibration by genus
$g$ curves given by $\pi: Y \rightarrow B$. Given an eight-form polynomial on
$Y$, $A_8$, we wish to be able to compute the integration over the fiber to a
six-form on the base, $B$.
\begin{equation}
  A_6 = \int_{\text{fiber}} A_8 = \pi_*A_8 \,.
\end{equation}
The $A_8$ that we shall be interested in will be those involving only Chern
classes on $Y$ or Chern classes of bundles over $B$ pulledback to $Y$. As such
we must determine how to integrate over the fiber, or pushforward, products of
Chern classes of $Y$ to forms on $B$.

Let us assume that the elliptic fibration is embedded as a (possibly singular)
hypersurface in an ambient projective space $X = \mathbb{P}(\mathcal{O}_B \oplus
\mathbb{L}^2 \oplus \mathbb{L}^3)$, and a projection onto the base $\rho: X
\rightarrow B$. This involves no loss of generality as any
elliptic fibration can be written as a Weierstrass model \cite{MR0387292},
where $\wL$ is part of the defining data known as the Weierstrass line bundle
\cite{MR1078016}. 
Furthermore we restrict ourselves to Weierstrass models which admit a
crepant resolution where the fibration is flat, i.e.~the fiber over each
point is complex one-dimensional.
Let $H =
c_1(\mathcal{O}_X(1))$ be the
hyperplane class of the projective fibration, and the elliptic fibration, $Y$, be
a hypersurface cut out by a polynomial of degree
\begin{equation}
  [Y] = 3 H + 6 \rho^* c_1(\mathbb{L}) \,,
\end{equation}
via the inclusion map $i : Y \hookrightarrow X$. The adjunction formula allows
one to determine the images of the Chern classes of $Y$ under this inclusion
map, as
\begin{equation}
  i_* c(Y) = \frac{(1 + H)(1 + H + 2 \rho^*c_1(\mathbb{L}))(1 + H + 3
    \rho^*c_1(\mathbb{L}))}{1 + 3 H + 6 \rho^*c_1(\mathbb{L})} \rho^*c(B) \,.
\end{equation}
Now we know that
\begin{equation}
  \pi_* c(Y) = \rho_* \left( i_* c(Y) [Y] \right) \,,
\end{equation}
and thus
\begin{equation}\label{eqn:pushy}
  \pi_* c(Y) = \rho_* \left( \frac{(1 + H)(1 + H + 2 \rho^*c_1(\mathbb{L}))(1 + H + 3
    \rho^*c_1(\mathbb{L}))}{1 + 3 H + 6 \rho^*c_1(\mathbb{L})} (3 H + 6
    \rho^*c_1(\mathbb{L}))\right) c(B)
    \,.
\end{equation}
We note that this result relies on the fact that $Y$ is a smooth hypersurface
-- this essentially restricts us to considering smooth Weierstrass models. In
fact if $Y$ is a singular Weierstrass model then we want to compute the
pushforward of the Chern classes of a crepant resolution of $Y$,
$\widetilde{Y}$. When $\widetilde{Y}$ is obtained from $Y$ by a blowup of
smooth centers $Z_1 = \cdots = Z_m$ one can use the results of \cite{MR2600139} to
augment (\ref{eqn:pushy}) with additional classes related to $Z_i$ and any
exceptional divisors introduced in the blowups. This modification does not
alter the $\mathbb{L}$-universal terms in the pushforward of the Chern classes of the
smooth fibration from those terms that proceed from (\ref{eqn:pushy})
-- therefore we only consider the contributions from (\ref{eqn:pushy}) to
$\pi_* c(\widetilde{Y})$.

We can compute the pushforward on the RHS of (\ref{eqn:pushy}) by taking
advantage of the intersection ring relation; in the Chow ring of $X$ we have
the generating relation
\begin{equation}
  H^3 = - 5 H^2 \rho^* c_1(\mathbb{L}) - 6 H \rho^* c_1(\mathbb{L})^2 \,, 
\end{equation} 
and that under the pushforward
\begin{equation}
  \rho_* H = 0 \,, \quad \rho_* H^2 = 1 \,. 
\end{equation}
To determine the term in the pushforward of $c_i(Y)$ proportional to the
highest degree of $c(B)$ we must determine the lowest non-trivial contribution
to the pushforward on the RHS of (\ref{eqn:pushy}). This is
\begin{equation}
    \rho_* \left( 12 H^2 \rho^* c_1(\mathbb{L}) + 60 H \rho^*
    c_1(\mathbb{L})^2 + 72 \rho^* c_1(\mathbb{L})^3 \right) = 12 
    c_1(\mathbb{L}) \,. 
\end{equation}
As such, one finds that
\begin{equation}\label{eqn:pushchern}
  \pi_* c_i(Y) = 12 c_1(\mathbb{L}) c_{i-2}(B) + \cdots \,,
\end{equation}
where $\cdots$ represents higher degree terms. These will depend on the
particular fibration $Y$, however the terms that are written explicitly in
(\ref{eqn:pushchern}) are $\mathbb{L}$-universal, for every elliptic fibration $Y$.

Using that the canonical bundle of an elliptic fibration is
a pullback bundle from the base, 
\begin{equation}
  c_1(Y) = \pi^*\left( c_1(B) - c_1(\wL) \right) \,,
\end{equation}
one can use (\ref{eqn:pushchern}) to compute the $\mathbb{L}$-universal contribution to
almost all of the degree four polynomials in the Chern classes of an elliptic
fourfold. These are
\begin{equation}\label{eqn:cnmost2}
  \begin{aligned}
    \pi_* \left(c_1(Y)^4\right) &= 0 \cr
    \pi_* \left(c_2(Y)c_1(Y)^2\right) &= 12 c_1(\wL) c_1(B)^2 + \cdots \cr
    \pi_* \left(c_3(Y)c_1(Y)\right) &= 12 c_1(\wL)c_1(B)^2 + \cdots \cr
    \pi_* \left(c_4(Y)\right) &= 12 c_1(\wL) c_2(B) + \cdots \,,
  \end{aligned}
\end{equation}
where the $\cdots$ represents terms arising from the non-universal terms,
$\cdots$, in (\ref{eqn:pushchern}).
There is a final Chern number for an elliptic fourfold which we can now
  compute using the pushforwards (\ref{eqn:cnmost2}) and the
  Grothendieck--Riemann--Roch theorem. GRR states that for any proper morphism
  between smooth varieties $\rho: X \rightarrow Z$, and for a bundle
  $\mathcal{E}$ on $X$ we have
\begin{equation}
  \rho_*\left( \ch(\mathcal{E}) \td(X) \right) = \sum_i (-1)^i
        \ch\left(R^i\rho_*\mathcal{E}\right)\td(Z) \,.
\end{equation}
For an elliptic fibration such that we consider it is known that
\begin{equation}
  R^1\pi_* \mathcal{O}_Y = \wL^\vee \,,
\end{equation}
and thus
\begin{equation}\label{eqn:GRR}
  \pi_* \td({Y}) = (1 - \ch(\wL^\vee))\td(B) \,.
\end{equation}
Using the three-form part of (\ref{eqn:GRR}) and the pushforwards
(\ref{eqn:cnmost2}) we can compute the final degree four polynomial
\begin{equation}\label{eqn:c2c2}
  \pi_*\left(c_2(Y)^2\right) = 24 c_1(\wL) c_2(B) + \cdots \,.
\end{equation}
Of interest is also the pushforwards of degree four monomials where a part
of the monomial is a form pulled back from the base of the fibration. A 
particularly common example is when $\alpha$ is a degree two form on $B$ and
then
\begin{equation}
  \pi_* \left( c_2(Y) \pi^*\alpha \right) = 12 c_1(\wL) \alpha \,,
\end{equation}
as can be read off directly from (\ref{eqn:pushchern}) and the projection
formula.

\section{Anomaly for $T[T^2, \mathcal{F}; U(1)]$ on $\Sigma^\tau$}

In \cite{Lawrie:2016axq} the spectrum of a single D3-brane topologically
duality twisted and compactified on a complex curve, $\Sigma$, is obtained. In
this appendix we summarise the construction of the anomaly polynomials from
the determined spectrum, for $\Sigma$ a holomorphic curve inside the K\"ahler
base, $B$, of an elliptically fibered Calabi--Yau threefold, fourfold, or a K3
surface.

In each case the spectrum was found to have two distinct sectors, a bulk
sector and a defect sectors. The bulk sector involved the zero-modes of the
fields in the abelian $\mathcal{N} = 4$ vector multiplet; the defect modes
correspond to the transverse $7$-branes which intersect the compactification
curve $\Sigma$ at points -- these defect modes are sometimes referred to as
$3$--$7$ strings. It was determined \cite{Lawrie:2016axq} from studying the
spectrum on a single M5-brane, in an extension of \cite{Vafa:1997gr}, that
such modes always contribute
\begin{equation}
  n_{37} = 8 c_1(B) \cdot \Sigma \,,
\end{equation}
left-moving Fermi multiplets. As the defect contribution is universal, we will
summarise the anomaly generated solely by the bulk spectrum in this appendix.

\subsection{$\Sigma^\tau \subset CY_3$}\label{app:CY3spec}

  As described in section \ref{sec:from4dspec}, $\mathcal{N} = 4$ SYM contains
  inside the vector multiplet Majorana--Weyl fermions transforming in the 
  \begin{equation}
    ({\bf 2,1,4})_{1/2} \oplus ({\bf 1,2,\bar{4}})_{-1/2} \,,
  \end{equation}
  of the $SO(1,3) \times SU(4)_R \times U(1)_D$ symmetry group.  Under the
  decomposition and duality twist for a curve $\Sigma \subset CY_3$,  the
  resulting 2d theory has an $SU(2)_+ \times SU(2)_-$ global symmetry, the
  first factor of which is the $(0,4)$ R-symmetry and the $SU(2)_-$ is an
  additional flavor symmetry.  There are four different kinds of fermions in
  the theory, with representations
\begin{equation}
  (\textbf{2,1})_+ \,, \quad (\textbf{2,1})_- \,, \quad (\textbf{1,2})_+ \,,
  \quad (\textbf{1,2})_+ \,,
\end{equation}
where the subscript is the 2d chirality, and where the number of zero-modes is
counted respectively by
\begin{equation}
  z_1 \,, \quad z_2 \,, \quad z_3 \,, \quad z_4 \,.
\end{equation}
The values of the $z_i$ for the topological duality twist were worked out in
\cite{Lawrie:2016axq}. For each fermion of each distinct kind the contribution
to the anomaly is given by
\begin{equation}
  \begin{aligned}
    ({\bf 2,1})_\pm &\rightarrow \pm \ch(N^+) \widehat{A}(M_2)|_{4\text{-form}} \cr
    ({\bf 1,2})_\pm &\rightarrow \pm \ch(N^-) \widehat{A}(M_2)|_{4\text{-form}} \,,
  \end{aligned}
\end{equation}
where $N^\pm$ are the vector bundles associated to the fundamental
representations of the two $SU(2)$ factors, exactly as also discussed in
section \ref{sec:CY3bdl}. In this manner, and using the relation between Chern
classes and Chern characters for $N^\pm$ that
\begin{equation}
  \ch(N^\pm) = 2 - c_2(N^\pm) \,,
\end{equation}
one finds that the total anomaly polynomial for this 2d theory is
\begin{equation}
  I_4 = 2(z_1 - z_2 + z_3 - z_4)\left( - \frac{1}{24} p_1(M_2) \right) - (z_1
  - z_2)c_2(N^+) - (z_3 - z_4)c_2(N^-) \,.
\end{equation}
In \cite{Lawrie:2016axq} the multiplicities of the fermions were determined to
be
\begin{equation}
    z_1 = g - 1 + c_1(B) \cdot \Sigma \,, \quad 
    z_2 = 0 \,, \quad
    z_3 = 1 \,, \quad
    z_4 = g \,,
\end{equation}
and thus the anomaly polynomial can be written as
\begin{equation}
  \begin{aligned}
    I_4 &= c_2(N^-) \left( \frac{1}{2} \Sigma \cdot \Sigma - \frac{1}{2} c_1(B)
    \cdot \Sigma \right)  + c_2(N^+)  \left( - \frac{1}{2} \Sigma \cdot \Sigma -
    \frac{1}{2} c_1(B)\cdot \Sigma \right) \cr &\quad - \frac{1}{24} 2 c_1(B)
    \cdot \Sigma p_1(TM_2) \,.
  \end{aligned}
\end{equation}
This is identical to the result from integrating the anomaly polynomial
(\ref{eqn:POLY}), as given in (\ref{eqn:I4conc}), for $N = 1$. Indeed if one
is sufficiently careful about factors of $i$ this also agrees with the anomaly
polynomial generated by the spectrum\footnote{Recall the previously mentioned
caveat about the $3$--$7$ strings.} of the compactification as given in
(5.28) of \cite{Lawrie:2016axq}.

\subsection{$\Sigma^\tau \subset CY_4$}\label{app:CY4spec}

In the strings with $(0,2)$ worldvolume supersymmetry arising in
\cite{Lawrie:2016axq} via a single D3-brane wrapping a curve in the base of an
elliptic Calabi--Yau fourfold. Such a theory has an $SO(2)_T$ flavor symmetry
corresponding to the rotations of the transverse non-compact space to the
string. The bulk spectrum of the worldvolume theory was computed using the
topological duality twist and found to be
\begin{equation}
  \begin{array}{c|c|c} 
    \text{Fermions} & SO(2)_T \text{ Charge} & \text{Multiplicity} \cr\hline
    \mu_+ & +1 & z_1 = h^0(\Sigma, N_{\Sigma/B}) \cr
    \psi_+ & -1 & z_2 = g - 1 + c_1(B) \cdot \Sigma \cr
    \gamma_+ & -1 & z_3 = 1 \cr
    \rho_- & +1 & z_4 = h^0(\Sigma, N_{\Sigma/B}) - c_1(B) \cdot \Sigma \cr
    \beta_- & -1 & z_5 = g \cr
    \lambda_- & -1 & z_6 = 0
  \end{array} \,.
\end{equation}
Let us introduce the rank one vector bundle $N$ which corresponds to the
$SO(2)_T$ representation with charge $+1$. The anomaly polynomial induced by
such a spectrum is thus
\begin{equation}
  \begin{aligned}
    I_4 &= \left( \left[ (z_1 - z_4) \ch(N) + (z_2 + z_3 - z_5 - z_6)
    \ch(\overline{N}) \right] \widehat{A}(M_2) \right)_{4\text{-form}} \cr
    &= \left( \frac{1}{2}c_1(N)^2 - \frac{1}{24} p_1(M_2) \right)\left( z_1 +
    z_2 + z_3 - z_4 - z_5 - z_6\right) \cr
    &= \frac{1}{2} \left( 2 c_1(B) \cdot \Sigma \right) c_1(N)^2 - \frac{1}{24}
    \left(2 c_1(B) \cdot \Sigma\right) p_1(M_2) \,.
  \end{aligned}
\end{equation}
Thus the 't Hooft anomaly coefficients for the bulk spectrum\footnote{Again,
not including the $3$--$7$ strings.} are
\begin{equation}
  \begin{aligned}
    k_{TT} &= 2 c_1(B) \cdot \Sigma \cr
    k = c_R - c_L &= 2 c_1(B) \cdot \Sigma \,.
  \end{aligned}
\end{equation}
We can see that this is the specialised $N = 1$ values of the 't Hooft
anomalies as worked out from integrating the anomaly polynomial
(\ref{eqn:POLY}) over the curve given in (\ref{eqn:CY4tHooft}).

\subsection{$\Sigma^\tau \subset K3$ }\label{app:K3spec}

When $U(1)$ $\mathcal{N} = 4$ SYM is compactified on the rational curve that
is the base of a K3 elliptic fibration, with $\tau$ varying along the curve,
the result is a 2d theory with $(0,8)$ supersymmetry, as described in
\cite{Lawrie:2016axq}. In such a case the $3$--$3$ spectrum, which does not
include the defect modes induced by the varying coupling, consists of a single 
$(0,8)$ hypermultiplet, and the theory has an $SO(6)$ flavor symmetry arising
from the transverse rotations to the string in the 8d space-time. The one-loop
anomaly polynomial induced from the chiral fermions in this multiplet is
simply
\begin{equation}
    I_4 = \left( \ch(\mathcal{S}_6^+) \widehat{A}(M_2) \right)_{4\text{-form}} 
    = - c_2(\mathcal{S}_6^+) - \frac{1}{24} 4 p_1(M_2) \,.
\end{equation}




\providecommand{\href}[2]{#2}\begingroup\raggedright\endgroup

\end{document}